\newcommand{\apj}{ApJ}

\newcommand{\mnras}{MNRAS}

\newcommand{\etal}{{et al.~}}

\documentclass[useAMS,usenatbib]{mn2e}
\usepackage{amssymb}

\usepackage{amsmath}

\usepackage[dvips]{graphicx}
\title[Modeling Galaxy-Galaxy Weak Lensing with SDSS Groups]
      {Modeling Galaxy-Galaxy Weak Lensing with SDSS Groups}

\author[Ran Li et. al]
       {\parbox[t]{\textwidth}{
        Ran Li$^{1,2}$\thanks{E-mail:ranl@astro.umass.edu},
        H.J. Mo$^{2}$,
        Zuhui Fan$^{1}$,
        Marcello Cacciato$^{3}$,
        Frank C. van den Bosch$^{3}$,\\
        Xiaohu Yang$^{4}$,
        Surhud More$^{3}$}
        \vspace*{3pt} \\
  $^{1}$Department of Astronomy, Peking University, Beijing 100871, China\\
  $^{2}$Department of Astronomy,  University of Massachusetts, Amherst
        MA 01003, USA \\
  $^{3}$Max-Planck Institute for Astronomy, K\"onigstuhl 17, D-69117
        Heidelberg, Germany\\
  $^{4}$Shanghai Astronomical Observatory, the Partner Group of MPA,
        Nandan Road 80, Shanghai 200030, China}
\begin{document}

\date{}
\pagerange{\pageref{firstpage}--\pageref{lastpage}}
\pubyear{2008}

\maketitle

\label{firstpage}


\begin{abstract}
  We  use galaxy  groups selected from the  Sloan Digital  Sky Survey
  (SDSS) together with mass models  for individual groups to study the
  galaxy-galaxy  lensing signals expected  from galaxies  of different
  luminosities  and   morphological  types.   We   compare  our  model
  predictions with the observational results obtained from the SDSS by
  Mandelbaum et  al.  (2006)  for the same  samples of  galaxies.  The
  observational results  are well  reproduced in a  $\Lambda$CDM model
  based on the WMAP 3-year  data, but a $\Lambda$CDM model with higher
  $\sigma_8$,  such  as  the  one  based  on  the  WMAP  1-year  data,
  significantly  over-predicts the  galaxy-galaxy lensing  signal.  We
  model,  separately, the contributions  to the  galaxy-galaxy lensing
  signals   from  different   galaxies:   central  versus   satellite,
  early-type  versus late-type,  and galaxies  in haloes  of different
  masses.   We also  examine how  the predicted  galaxy-galaxy lensing
  signal depends  on the shape,  density profile, and the  location of
  the central galaxy with respect to its host halo.
\end{abstract}


\section{Introduction}

According  to the  current paradigm  of structure  formation, galaxies
form and reside inside extended cold dark haloes.  While the formation
and evolution  of dark  matter haloes in  the cosmic density  field is
mainly  determined  by  gravitational  processes,  the  formation  and
evolution  of  galaxies involves  much  more  complicated, and  poorly
understood processes,  such as radiative cooling,  star formation, and
all  kinds  of feedback.   One  important  step  in understanding  how
galaxies form and  evolve in the cosmic density  field is therefore to
understand how  the galaxies  of different physical  properties occupy
dark matter haloes of different masses.  Theoretically, the connection
between galaxies and dark matter haloes can be studied using numerical
simulations (e.g.,  Katz, Weinberg  \& Hernquist 1996;  Pearce et  al.
2000; Springel 2005; Springel  et al.  2005) or semi-analytical models
(e.g.  White \&  Frenk 1991; Kauffmann et al.   1993, 2004; Somerville
\& Primack 1999;  Cole et al.  2000;  van den Bosch 2002; Kang  et al.
2005; Croton et al.  2006).  These approaches try to model the process
of  galaxy  formation  from  first  principles.   However,  since  our
understanding of  the relevant processes is still  poor, the predicted
connection between  the properties of galaxies and  dark matter haloes
needs  to be  tested against  observations.  More  recently,  the halo
occupation model  has opened another  avenue to probe  the galaxy-dark
matter halo  connection (e.g.  Jing,  Mo \& B\"orner 1998;  Peacock \&
Smith 2000; Berlind  \& Weinberg 2002; Cooray \&  Sheth 2002; Scranton
2003; Yang, Mo \& van den Bosch  2003; van den Bosch, Yang \& Mo 2003;
Yan, Madgwick \&  White 2003; Tinker et al.  2005;  Zheng et al. 2005;
Cooray 2006; Vale \& Ostriker 2006; van den Bosch et al.  2007).  This
technique uses the observed  galaxy luminosity function and clustering
properties  to  constrain the  average  number  of  galaxies of  given
properties that occupy a dark matter halo of given mass.  Although the
method has  the advantage that  it can yield  much better fits  to the
data  than the  semi-analytical models  or numerical  simulations, one
typically  needs  to  assume  a  somewhat ad-hoc  functional  form  to
describe the halo occupation model.

A more  direct way  of studying the  galaxy-halo connection is  to
use galaxy groups\footnote{In this paper, we refer to a system of
galaxies as a group regardless of its richness, including isolated
galaxies (i.e., groups with a single member) and rich clusters of
galaxies.}, provided that they are defined as sets of galaxies that
reside in  the same dark matter  halo.  Recently, Yang et al. (2005;
2007) have developed  a halo-based group finder that is optimized
for grouping galaxies that reside in the same dark matter halo.
Using mock galaxy  redshift surveys constructed  from the
conditional luminosity function  model  (e.g.   Yang et al. 2003)
and a semi-analytical model (Kang et al. 2005), it is found that
this group finder is  very successful in associating galaxies with
their common dark matter haloes (see Yang et al. 2007; hereafter
Y07). The group finder  also performs reliably for poor systems,
including isolated galaxies in small mass haloes, making it ideally
suited for the study of the relationship between galaxies and dark
matter haloes  over a wide range  of  halo masses. However, in order
to interpret  the properties of the galaxy systems in terms of dark
matter haloes, one needs to  know the halo mass associated with each
of the groups. One approach commonly adopted is to  use some halo
mass indicator (such as the total stellar mass or luminosity
contained in member galaxies) to rank the groups.   With the
assumption that the  corresponding halo masses have the same ranking
and  that the mass function of the haloes associated  with groups is
the same as that given  by  a model  of structure formation, one can
assign  a halo  mass  to each of  the observed groups. This approach
was adopted by  Y07 for the  group catalogue used in this paper.
There are three potential problems with this approach. First, the
approach is model-dependent, in  the sense that the assumption  of a
different model of  structure formation will lead to  a different
halo mass function,  and hence  assign different masses to the
groups. Second, even if the assumed model of structure formation is
correct, it  is still  not  guaranteed that the  mass assignment
based on the ranking  of group stellar mass (or luminosity) is
valid. Finally, even if  all groups are assigned with accurate halo
masses, the question how dark  matter is distributed within the
galaxy groups remains open. Clearly,  it is important to have
independent mass measurements of the haloes associated with galaxy
groups to test the validity of the mass estimates based on
the stellar mass (luminosity) ranking.

Gravitational   lensing   observations,   which  measure   the image
distortions of  background galaxies caused by  the gravitational
field of the matter distribution in the foreground, provide a
promising tool to  probe  the  dark  matter distribution directly.
In  particular, galaxy-galaxy  weak lensing,  which focuses  on the
image distortions around lensing galaxies, can be used to probe the
distribution of dark matter  around   galaxies, hence  their  dark
matter  haloes.   The galaxy-galaxy  lensing signal  produced  by
individual  galaxies  is usually very weak, and so one has to stack
the  signal from many lens galaxies  to have a statistical
measurement.  The  first  attempt to detect such galaxy-galaxy
lensing signal  was reported by Tyson et al. (1984). More recently,
with the  advent of  wide and  deep surveys, galaxy-galaxy lensing
can be studied  for lens galaxies  of different luminosities,
stellar masses,  colors and  morphological types  (e.g. Brainerd et
al. 1996; Hudson et al. 1998; McKay et al. 2001; Hoekstra et  al.
2003; Hoekstra 2004;  Sheldon et  al.  2004; Mandelbaum et  al.
2005, 2006; Sheldon et  al.  2007a;  Johnston et al. 2007;  Sheldon
et al. 2007b; Mandelbaum et  al. 2008).  Given that galaxies reside
in dark matter haloes,  these  results provide  important
constraints on the mass distribution associated with galaxies in a
statistical way.

In this paper, we use the galaxy groups of Y07 selected from the Sloan
Digital Sky  Survey (SDSS), together  with mass models  for individual
groups, to predict the galaxy-galaxy lensing signal expected from SDSS
galaxies.   We compare  our model  predictions with  the observational
results obtained by  Mandelbaum et al. (2006) for  the same galaxies.
Our goal  is threefold. First, we  want to test whether  the method of
halo-mass assignment to  groups adopted by Y07 is  reliable. Since the
method provides a  potentially powerful way to obtain  the halo masses
associated  with the  galaxy  groups, the  test  results have  general
implications for  the study of  the relationship between  galaxies and
dark  matter  haloes.   Second,  we  want to  examine  in  detail  the
contributions  to  the  galaxy-galaxy  lensing signal  from  different
systems, such as central  versus satellite galaxies, early-type versus
late-type galaxies, and groups  of different masses. Such analysis can
help us interpreting the observational results. Finally, we would like
to  study how the  predicted galaxy-galaxy  lensing signal  depends on
model  assumptions, such  as the  cosmological model  and  the density
profiles of dark matter haloes. In  a companion paper (Cacciato et al.
2008,  hereafter C08), we  use the  relationship between  galaxies and
dark matter  haloes obtained from the  conditional luminosity function
(CLF)  modeling (Yang  et al.   2003; van  den Bosch  et al.  2007) to
predict  the  galaxy-galaxy cross  correlation  and  to calculate  the
lensing signal, while here we  directly use the observed galaxy groups
and their galaxy memberships.

This paper  is organized as follows. In  Section \ref{sec_analysis}
we define  the statistical measure  that characterizes  the
galaxy-galaxy lensing effect expected from the mass distribution
associated with the galaxy groups.   We provide  a brief description
of the  galaxy group catalogue  and the  models of  the mass
distribution  associated with galaxy  groups in Section
\ref{sec_SDSS}. We  present our  results in Section
\ref{sec_results}      and     conclude     in     Section
\ref{sec_conclusions}.   Unless   specified  otherwise,  we   adopt
a $\Lambda$CDM cosmology  with parameters given by the  WMAP 3-year
data (Spergel  et al.  2007,  hereafter WMAP3  cosmology) in  our
analysis: $\Omega_{\rm m}=0.238$,  $\Omega_{\Lambda}=0.762$, and
$h\equiv H_0/(100 \rm {km\,s^{-1} \,Mpc^{-1}})=0.73$,
$\sigma_8=0.75$ .

\section{Galaxy-Galaxy Lensing}
\label{sec_analysis}

Galaxy-galaxy lensing provides a statistical measure of the profile of
the tangential  shear, $\gamma_t(R)$, averaged over a  thin annulus at
the projected  radius $R$ around  the lens galaxies. This  quantity is
related to the excess surface  density (hereafter ESD) around the lens
galaxy, $\Delta\Sigma$, as
\begin{equation}
\Delta\Sigma(R)=\gamma_t(R)\Sigma_{\rm crit}=\bar{\Sigma}(<R)-\Sigma(R)\,,
\end{equation}
where $\bar{\Sigma}(<R)$  is the  average surface mass  density within
$R$, and  $\Sigma(R)$ is the  azimuthally averaged surface  density at
$R$.   Note that,  according  to this  relation, $\Delta\Sigma(R)$  is
independent of a uniform background. In the above equation,
\begin{equation}
\Sigma_{\rm crit}=\frac{c^2}{4\pi G}\frac{D_s}{D_l D_{ls}(1+z_l)^2}
\end{equation}
is the  critical surface  density in comoving coordinates, with
$D_s$ and $D_l$ the angular distances  of the  lens  and source,
$D_{ls}$ the angular distance between the source and the lens, and
$z_l$ the redshift of the lens.

By  definition, the surface  density, $\Sigma(R)$,  is related  to the
projection of the  galaxy-matter cross-correlation function, $\xi_{\rm
  g,  m}(r)$,  along  the   line-of-sight.  In  the  distant  observer
approximation
\begin{equation}\label{xi_gm}
\Sigma(R)=\bar{\rho}\int ^{\infty}_{-\infty}\left
[1+\xi_{\rm g,m}(\sqrt{R^2+\chi^2}) \right ]\,d\chi\, ,
\end{equation}
where $\bar{\rho}$ is  the mean density of the  universe and $\chi$ is
the line-of-sight  distance from the  lens.

The  cross-correlation  between  galaxies  and  dark  matter  can,
in general, be divided  into a 1-halo term and a  2-halo term. The
1-halo term measures  the cross-correlation between galaxies  and
dark matter particles in their own host haloes, while the 2-halo
term measures the cross-correlation between galaxies and  dark
matter particles in other haloes.  In the present work, we are
interested in the lensing signals on  scales  $R  \leq   2
h^{-1}{\rm  Mpc}$  where  the  observational measurements   are  the
most   accurate.   As   we   will  show   in \S~\ref{sec_results},
on such scales the signal is mainly dominated by the 1-halo  term.
Nevertheless, our model also  takes the contribution of  the  2-halo
term  into account. More importantly,  since  central galaxies
(those residing at  the center  of a  dark matter  halo) and
satellite galaxies (those orbiting around a central galaxy)
contribute very different lensing signals  (e.g.  Natarajan, Kneib
\& Smail 2002; Yang \etal  2006; Limousin \etal 2007),  it is
important  to model the contributions from central and satellite
galaxies separately.

\begin{figure}
\includegraphics[width=0.5\textwidth]{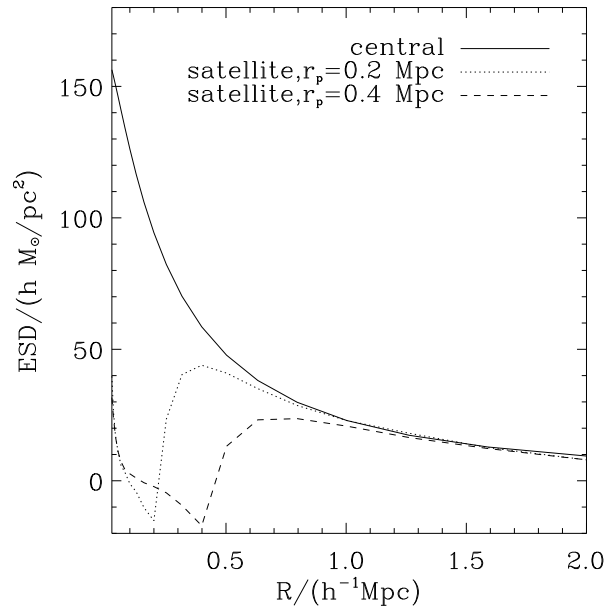}
\caption{The ESD expected for a single galaxy. Here the host halo
  mass is  assumed to  be $10^{14} h^{-1}M_{\odot}  $. The  solid line
  represents the lensing signal for the central galaxy in such a halo.
  The dotted line represents the  lensing signal of a satellite galaxy
  residing in  a sub-halo of  mass $10^{11} h^{-1}M_{\odot}$  which
  has a projected distance  $r_{\rm p}=0.2  h^{-1}{\rm Mpc}$ from  the center of
  the  host halo.  The dashed  line is  the same  as the  dotted line,
  except that the subhalo's projected distance is $0.4 h^{-1}{\rm Mpc}$ from
  the center of the host halo.}\label{fig:sig}
\end{figure}

As an illustration,  in Fig.  \ref{fig:sig} we show  the ESDs expected
from  a single  galaxy in  a host  halo of  mass $10^{14}  h^{-1} {\rm
  M}_{\odot}$.  The solid line  represents the lensing signal expected
for the central galaxy of the  halo. While the dotted and dashed
lines show the lensing  signal of a satellite galaxy  residing in a
sub-halo of $10^{11} h^{-1} {\rm M}_{\odot}$ with a projected
halo-centric distance $r_p=0.2\,h^{-1}{\rm  Mpc}$ and
$r_p=0.4\,h^{-1}{\rm  Mpc}$ from the center of the  host halo,
respectively.  In the calculation, the dark matter mass distribution
in  the host halo  is assumed to follow the Navarro, Frank \&  White
(1997) profile and that  in the sub-haloes is assumed to follow  the
Hayashi et al. (2003)  model. These models are described in detail
in \S~\ref{sec:distrib}. In order to estimate the ESD, we sample
these profiles with mass particles and project the positions of  all
particles  to a plane perpendicular to the  line of sight. The
$\Sigma(R)$ is then estimated by counting the number of dark matter
particles in a annulus with radius  $R$ centred on the selected
galaxies.  Fig. \ref{fig:sig} shows clearly that the lensing signals
of the central and the satellite are quite different.  The ESD of
the central galaxy follows the mass distribution  of the host halo,
decreasing monotonically with  $R$. The  ESD of  a satellite,  on
the other hand, consists of two parts: one from the subhalo
associated the satellite, which contributes to the inner part, and
the other from the host  halo,  which  dominates at  larger  $R$.
This  simple model demonstrates clearly that, in order to model the
galaxy-galaxy lensing signal produced by  a population of lens
galaxies,  one needs to model carefully  the distribution  of matter
around  both host haloes  and subhaloes. To do this,  we need not
only to identify the  haloes in which each lens galaxy resides, but
also to model the mass and density profile of  each host halo and
subhalo. In  addition we also need to model the  distribution of
dark matter relative to galaxies.   In the following section, we
describe our modeling with the  use of observed galaxy groups.

\section{Modeling the Mass Distribution Associated with the SDSS Groups}
\label{sec_SDSS}

\subsection{The SDSS Group Catalogue}

 Our analysis is  based on the SDSS galaxy group catalogue
constructed by  Y07.  The groups  are selected  with  the adaptive
halo-based group finder developed by  Yang et al. (2005), from the
New York University Value Added Galaxy Catalog (NYU-VAGC; Blanton et
al. 2005) which is based  on the SDSS Data Release 4
(Adelman-McCarthy et al. 2006). Only galaxies with redshifts in the
range $0.01\leq z \leq 0.2$, and  with redshift completeness
$\mathcal {C}>0.7$,  are used in the  group identification. The
magnitudes and colors of all galaxies are based on the standard SDSS
Petrosian technique (Petrosian 1976; Strauss et al. 2002), and have
been corrected for galactic extinction (Schlegel, Finkbeiner \&
Davis 1998). All magnitudes have been  K-corrected and
evolution-corrected to $z=0.1$ following  the method described in
Blanton et al. (2003). In Y07, three group samples were constructed
using galaxy samples of different sources of galaxy redshifts. Our
analysis is based on Sample II, which includes 362,356 galaxies with
redshifts from the SDSS and 7091 galaxies with redshifts taken  from
alternative surveys: 2dFGRS (Colless et al. 2001), PSOz (Saunders et
al.2000) or from the RC3 (de Vaucouleurs et al. 1991). There are  in
total 301,237 groups, including those with only one member galaxy.
The group finder has been applied to mock catalogue to test the
completeness and purity of the groups in Y07. About $90\%$ of the
groups have a completeness $f_c>0.6$ and $80\%$ groups with
$f_c>0.8$, where $f_c$ is defined as the ratio  between the number
of true members that are selected as the members of the group and
the number of the total true members of the group.

\subsection{Halo Mass Assignment}
\label{ssec_massassign}

\begin{figure*}
\includegraphics[width=1.0\textwidth]{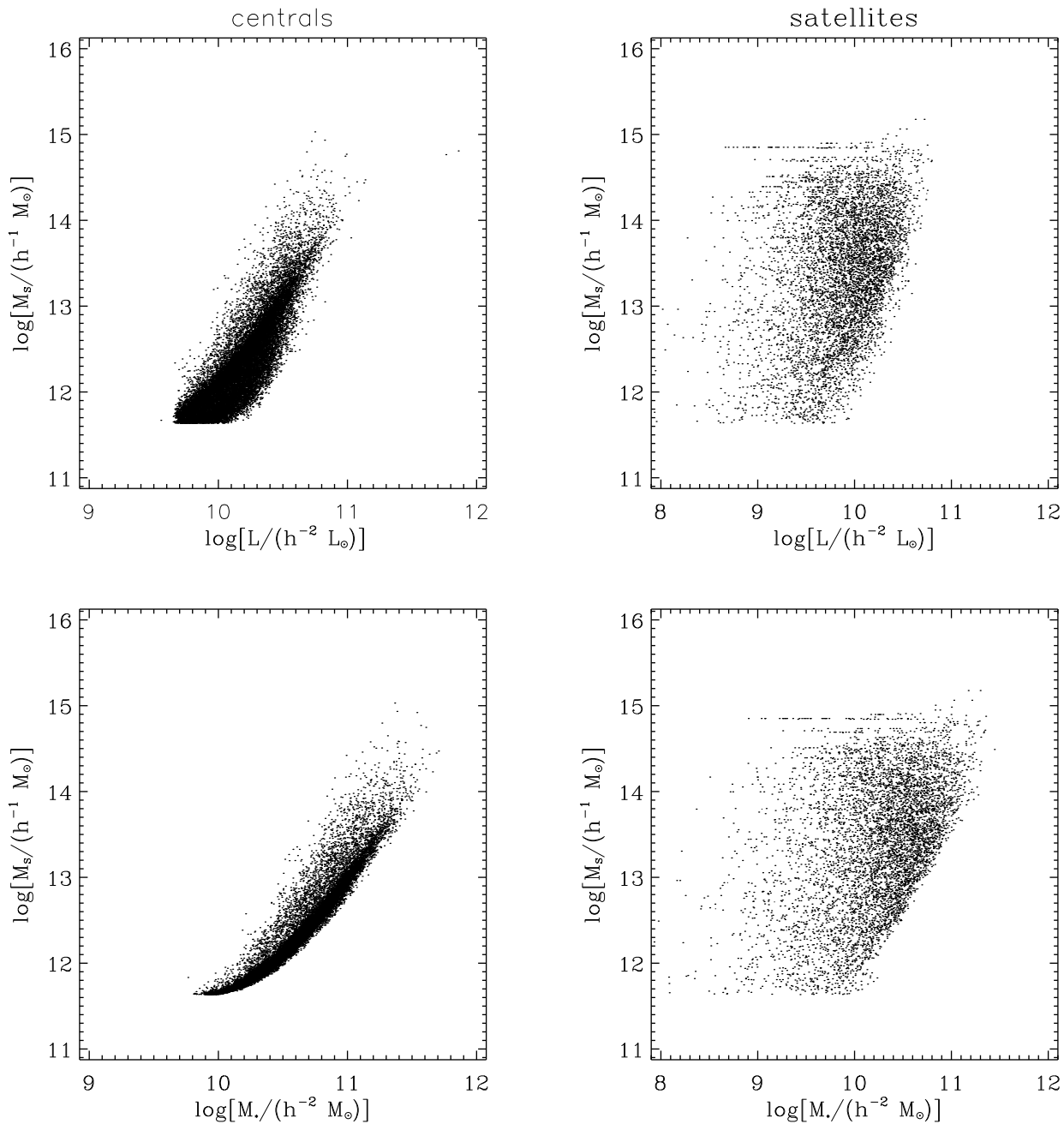}
\caption{The halo mass $M_S$ (estimated using stellar mass), versus
  $M_{\rm \ast}$ (lower panels) and $L$ (upper panels) of the galaxies
  in  the haloes. The  left panels  are for  central galaxies  and the
  right panels are for satellite galaxies.}\label{fig:ML}
\end{figure*}

\begin{figure*}
\includegraphics[width=1.0\textwidth]{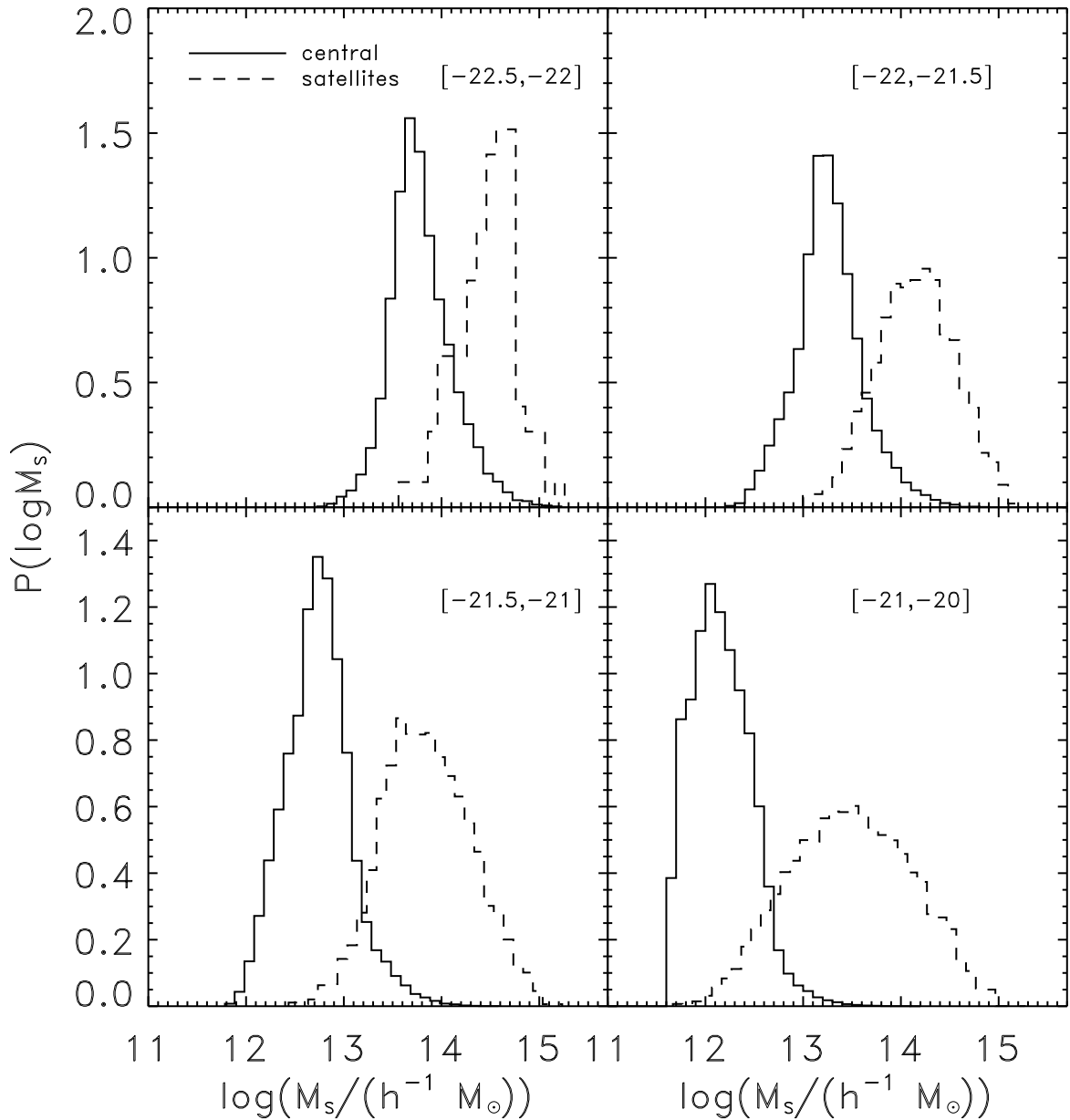}
\caption{The distribution of the host halo masses for the central
  and satellite galaxies in different luminosity bins, as indicated by
  the  $r$-band  absolute-magnitude  range in  each  panel.}\label{fig:dis}
\end{figure*}

An  important  aspect  of   the  group  catalog  construction  is  the
determination of the halo mass,  $M_{\rm vir}$, of each group. In Y07,
two estimators are  adopted. The first, $M_L$, is  estimated using the
ranking  of the  characteristic luminosity  of a  group, which  is the
total luminosity of all member galaxies in the group with $M_r-5\log h
\le -19.5$  (hereafter referred to as $L_{19.5}$).  The second, $M_S$,
is  estimated using the  ranking of  the characteristic  stellar mass,
$M_{\rm stellar}$  which is  defined to be  the total stellar  mass of
group members with $M_r-5\log h\le -19.5$. For each galaxy the stellar
mass  is estimated  from its  absolute magnitude  and color  using the
fitting formula given by Bell et al. (2003).

The  basic  assumption  of the  ranking  method  is  that there  is
a one-to-one relation between $M_{\rm  stellar}$ (or $L_{19.5}$) and
the group mass.  Using  the dark matter halo mass  function
predicted by a model of structure formation, one can assign a halo
mass to each group according  to  its  $M_{\rm  stellar}$  - ranking
(or  $L_{19.5}$  - ranking).  In this paper,  we use the mass
function obtained by Warren et al. (2006).
Note that this one-to-one
mapping is applicable only when the group sample is complete. In
Y07, three complete samples are constructed in three redshift
ranges. Only groups in the complete samples  are used in the
ranking. The mass of other groups  are estimated   by a linear
interpolation based on the $M_{\rm
  stellar}$-$M_{\rm vir}$ relation (or the $L_{19.5}$ - $M_{\rm vir}$)
obtained from  the complete sample.  Detailed tests  using mock galaxy
redshift samples have shown that the 1-$\sigma$ error of the estimated
halo  mass  is  $\sim  0.3$  dex  (Y07). In  addition,  the  two  mass
estimators, $M_L$  and $M_S$, agree  remarkably well with  each other,
with a scatter  that decreases from about 0.1 dex  at the low-mass end
to about 0.05 dex at  the high-mass end. Since the correlation between
$M_{\rm stellar}$ and halo mass  is somewhat tighter than that between
$L_{19.5}$ and  halo mass,  we adopt $M_S$  as our fiducial  halo mass
throughout. As we demonstrate in \S~\ref{sec:depend}, using $M_L$
instead yields results that are fairly similar.

Fig.~\ref{fig:ML}  shows  the relation  between  the  host halo
mass, $M_S$,  and the  galaxy stellar  mass $M_{\ast}$ (the  lower
two panels)  or the galaxy  luminosity  $L$ (the  upper  two
panels). Results  are  shown separately for  central galaxies (left
panels) and satellite galaxies (right  panels). As  one can  see,
the stellar  mass (luminosity)  of central  galaxies is  quite
tightly correlated with  their  host halo masses.  However, for
satellite galaxies of a given  stellar mass (or luminosity), their
host halo  mass covers a very large  range, reflecting the fact that
many low-mass galaxies are satellites  in massive  haloes. The
distributions of host halo masses,  $M_S$, for  central  or
satellite galaxies   in different   luminosity   bins  are   shown
in  Fig. \ref{fig:dis}. On  average, brighter central galaxies
reside in more massive haloes.  For  faint galaxies,  the halo-mass
distribution is broader, again  because many faint galaxies are
satellites in massive systems.

In the  group catalogue, the  mass assignment described above  is
used only for  groups where the brightest galaxy  is brighter than
$M_r - 5\log h = -19.5$. This is because the mass ranking used in
the group catalog is based on the total stellar mass (or total luminosity)
of the member galaxies that are brighter than $M_r - 5\log h = -19.5$.
The groups with no galaxies brighter than this magnitude thus have
no assigned rank. As described in Y07, the reason for choosing
this maginitude is a compromise between having a complete sample in
a relatively large volume and having more groups that are represented
by a number of member galaxies. For
groups in which all member galaxies have $M_r - 5\log h>-19.5$, a
different method has to be adopted. In modeling the luminosity
function and stellar mass function of the central galaxies based  on
the same SDSS group catalogue as used here,  Yang et  al. (2008)
obtain an average relation between the luminosity (or stellar mass)
of the central galaxy and the halo mass down to $M_r - 5\log h \sim
-17$. We adopt this relation to assign halo masses to all groups
(including those containing only one isolated galaxy) represented by
centrals with $M_r - 5\log h
>-19.5$. For convenience, the halo masses obtained in this  way are
also referred  to as  $M_S$ (based  on the stellar mass  of central
galaxies) and $M_L$  (based on  the $r$-band luminosity of the
central galaxies), respectively.

\subsection{Mass Distribution in Haloes and Subhaloes}
\label{sec:distrib}

With the group catalogue described above, we can model the dark
matter distribution  by convolving  the  halo distribution  with the
density profiles  of  individual  haloes.   In  our modeling  of the
density profiles, the  host halo of a group  is assumed to be
centered on the central galaxy. There are two ways  to define a
central galaxy: one is to define  the central in  a group to  be the
galaxy with  the highest stellar  mass, and  the  other is  to
define the  central  to be  the brightest member.  For most groups
these two definitions give the same results, but there are very few
cases (less than $\sim 2\%$) where different central galaxies are
defined. In our fiducial model, we define the most massive galaxies
(in term of stellar mass) to be the central galaxies.

In a hierarchical model, a dark  matter halo forms through a series
of merger events.  During the assembly of a halo, most of the mass
in the merging progenitors is expected to be stripped.  However,
some of them may  survive  as  subhaloes,  although  the total  mass
contained  in subhaloes is small, typically $\sim 10\%$ (van den
Bosch, Tormen \& Giocoli  2005). Some of  the subhaloes are
associated with `satellite galaxies'  in a halo. In our modeling of
the galaxy-galaxy weak lensing, we only take into  account subhaloes
associated  with satellite  galaxies, treating other subhaloes as
part of the host halo. Giocoli,  Tormen \& van den Bosch (2008)
provide  a fitting function of the  average mass function for
subhaloes at the time of their accretion into the parent halo of a
given mass. Using this mass function,  we first sample a set of
masses for  each group mass. We  then set  the mass  originally
associated  with a satellite  galaxy  according  to  the  stellar
mass  ranking  of  the satellites in  the group. Here we implicitly
assume that the initial subhalo mass function is the same as the
mass function of the subhalos that host satellites. This assumption
is not proved by any observations, and we have to live with it since
a more realistic model is not currently available. Fortunately,
subhalos only contribute a small fraction to the total lensing
signal on small scales. The uncertainty here will not have a
significant impact on any of our conclusions. To obtain the final
mass in the subhalo at the present time, the evolution of the
subhaloes needs to be taken into account. In other words, we need to
know the fraction of the mass that is stripped and how the structure
of a subhalo changes after the stripping. Here it is convenient  to
introduce a parameter $f_m$ which is  the retained mass fraction  of
the subhalo. Gao  et al. (2004) studied the radial dependence of the
retained mass fraction $f_m$ from a large sample of  subhaloes in a
large cosmological simulation. In their work, $f_m$ is considered as
a function of $r_s/r_{\rm vir, h}$, where $r_s$ is the distance of
the subhalo from the center of the host halo and $r_{\rm vir, h}$ is
the virial radius of  the host halo. The simulation of Gao et al.
gives
\begin{equation}\label{eq:gao}
f_m=0.65 (r_s/r_{\rm vir, h})^{2/3}\,.
\end{equation}
We  will adopt  this in  our modeling  of the  masses  associated
with subhaloes.  However,  in  the  group  catalogue,  only  the
projected distance, $r_p$, from the group center is available. The
3D-distance, $r_s$, is  obtained by randomly sampling the NFW
profile of the host halo with the given projected radius $r_p$.

Thus, the  mass assigned to a  subhalo is determined  by the
following three factors: (1)  the stellar mass of the satellite
galaxy; (2) the host halo mass; (3) the distance between the
satellite and the center of the host.  Here the host halo mass comes
into our calculation in two ways. It not only determines the
subhalo mass function, but also affects the parameter $f_m$ in
Eqs.\ref{eq:gao}.  Note that the accretion history of the host halo
may also affect the value of $f_m$. We have to neglect such effect
because it is unclear how to model the accretion histories for
individual groups.

For host haloes, we use the following NFW profile (Navarro, Frank \&
White 1997) to model the mass distribution:
\begin{equation}\label{eq:NFW}
\rho(r)=\frac{\delta_0 \bar{\rho}}{(r/r_c)(1+r/r_c)^{2}}\,.
\end{equation}
where $\bar{\rho}$  is the  mean density of  the universe, $r_c$  is a
scale  radius,   related  to  virial  radius  $r_{\rm   vir}$  by  the
concentration,   $c=   r_{\rm   vir}/r_c$,   and   $\delta_0$   is   a
characteristic over-density  related to the average  over-density of a
virialized halo, $\Delta_{\rm vir}$, by
\begin{equation}
\delta_0=\frac{\Delta_{\rm vir}}{3}\frac{c^3}{\ln(1+c) -c/(1+c)}\,.
\end{equation}
We  adopt the  value  of  $\Delta_{\rm vir}$  given  by the
spherical collapse  model (see Nakamura  \& Suto  1997; Henry 2000).
Numerical simulations  show that  halo concentrations  are
correlated  with halo mass,  and  we use  the  relations  given  by
Macci\`o  \etal  (2007), converted to our definition of halo mass.
Note that here we use $r_c$, instead of the conventional
notation $r_s$, to denote the scale radius
of the NFW profile, as  $r_s$ has been used to denote
the distance of a subhalo from the center of its host.

For  sub-haloes, we  model their  density profiles  using  the results
obtained by Hayashi et al. (2003), who found that the density profiles
of stripped sub-haloes can be approximated as
\begin{equation}\label{eq:Hayashi}
\rho_s(r)=\frac{f_t}{1+(r/r_{t, \rm eff})^3}\rho(r)\,,
\end{equation}
where $f_t$ is a dimensionless  factor describing the reduction in the
central density, and  $r_{t, \rm eff}$ is a  cut-off radius imposed by
the tidal force  of the host halo.  For  $f_t=1$ and $r_{t,\rm eff}\gg
r_c$, $\rho_s(r)$ reduces to  the standard NFW profile $\rho(r)$. Here
$\rho(r)$ is calculated  using the mass of the subhalo  at the time of
its accretion  into the  host halo.  Both  $f_t$ and $r_{t,  \rm eff}$
depend on the mass fraction of the sub-halo that remains bound, $f_m$.
Based on $N$-body simulations,  Hayashi et al.  obtained the following
fitting formulae relating $f_t$ and $r_{t,\rm eff}$ to $f_m$:
\begin{equation}
\log(r_{t, \rm eff}/r_c)=1.02+1.38\log f_m+0.37(\log f_m)^2\,;
\end{equation}
\begin{equation}
\log(f_t)=-0. 007+0.35\log f_m+0.39(\log f_m)^2+0.23(\log f_m)^3\,.
\end{equation}
It  should  be  pointed   out,  though,  that  there  are
substantial uncertainties  in  modeling the  mass  distribution
around  individual satellite galaxies. In particular, many of the
results about subhaloes are  obtained  from  $N$-body  simulations,
and  it  is  unclear  how significant the effect of  including
baryonic matter is.  Fortunately, the   total  mass   associated
with   satellite  galaxies   is  small (see   e.g.   Weinberg  et
al.   2008).  Furthermore,   the contribution of  the subhaloes
associated with  the satellite galaxies to the  galaxy-galaxy
lensing signal  is confined to small  scales. We therefore expect
that these  uncertainties will not change our results significantly.

With  the mass  distributions described  above, we  use  a
Monte-Carlo method  to sample  each of  the  profiles with  a random
set of  mass particles. Note  that the halo  mass assigned to  a
group in  the SDSS Group Catalog is $M_{180}$, which  is the mass
enclosed in the radius, $r_{180}$, defined such that $M_{180}= {4
  \pi}r_{180}^3(180\bar{\rho})/3$.  We therefore  sample  the particle
distribution within $r_{180}$. After all the particles in each halo
are sampled, we project the positions of all the particles to a
plane and calculate $\Delta\Sigma(R)$ by stacking galaxies in each
of the luminosity bin. Since the mass distribution is isotropic, an
arbitrary direction can be chosen for the stacking. Thus, the
projection effect is naturally included in our calculation. Each of
the particles has a mass of $10^{10}h^{-1}{\rm M}_{\odot}$. Our test
using particles of lower masses shows that the mass resolution
adopted here  is sufficient for our purpose. Using 2 times more
particles leads to a difference of about $5\%$ at $R \sim 0.02\,
h^{-1} {\rm Mpc}$, and almost no difference at $R \ge 0.1\,  h^{-1}
{\rm Mpc}$.

\begin{figure*}
\includegraphics[height=0.7\textheight, width=\textwidth]{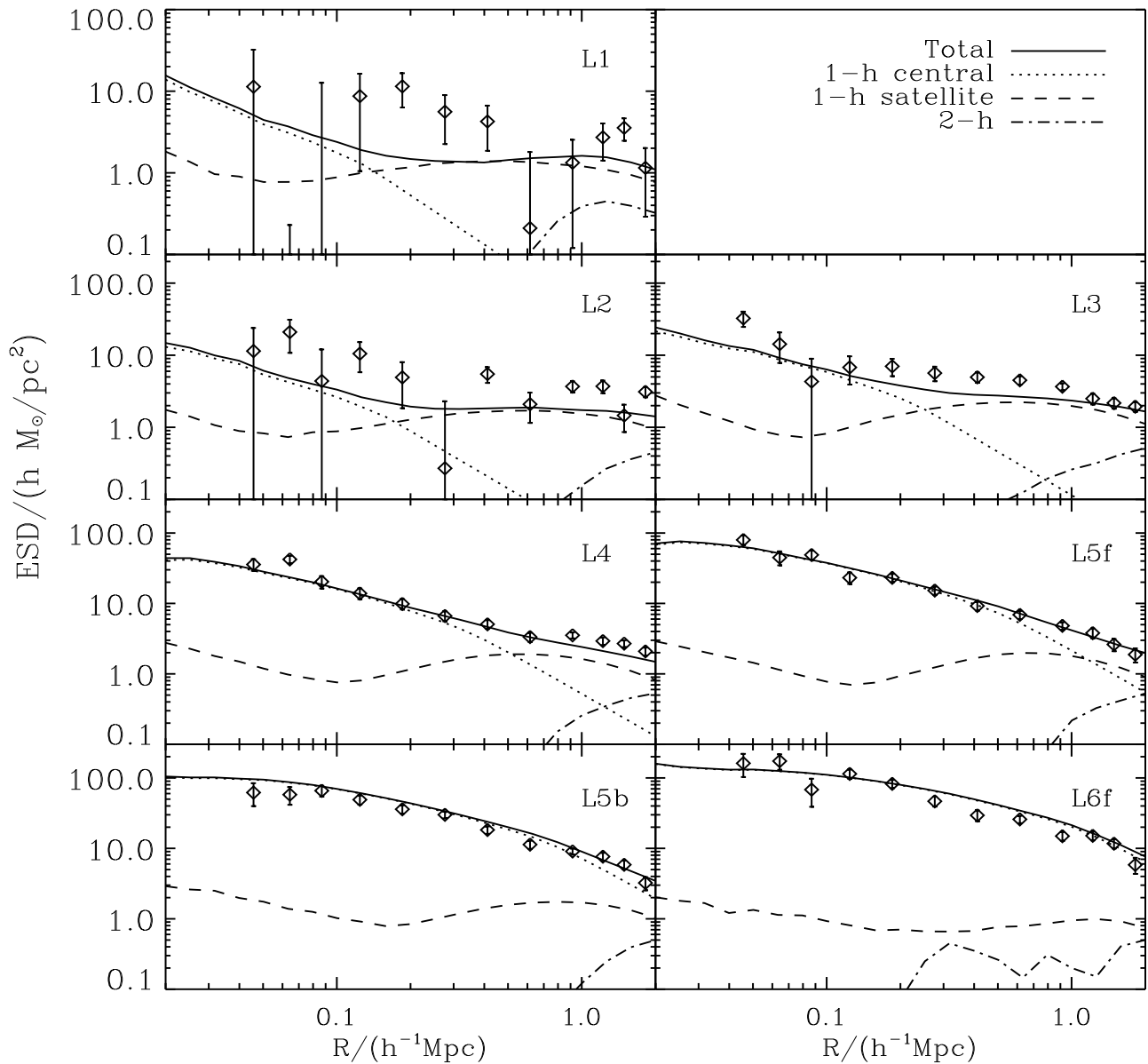}
\caption{Comparison of the lensing signal predicted by the fiducial
  model with the observational results. Here the ESD is plotted as the
  function  of the  transverse distance  $R$ for  lensing  galaxies in
  different  luminosity bins.  Data  points with  error  bars are  the
  observational results of Mandelbaum et al. , while the lines are the
  model predictions. The dotted, dashed and dot-dashed lines represent
  the  contributions of  the `1-halo  term' of  central  galaxies, the
  `1-halo term' of satellite galaxies,  and the `2-halo term' (of both
  centrals  and satellites),  respectively. The  solid lines  show the
  predicted total ESD. The $r$-band  magnitude range for each case can
  be found in Table 1.}\label{fig:4bs}
\end{figure*}

\begin{figure*}
\includegraphics[height=0.7\textheight, width=\textwidth]{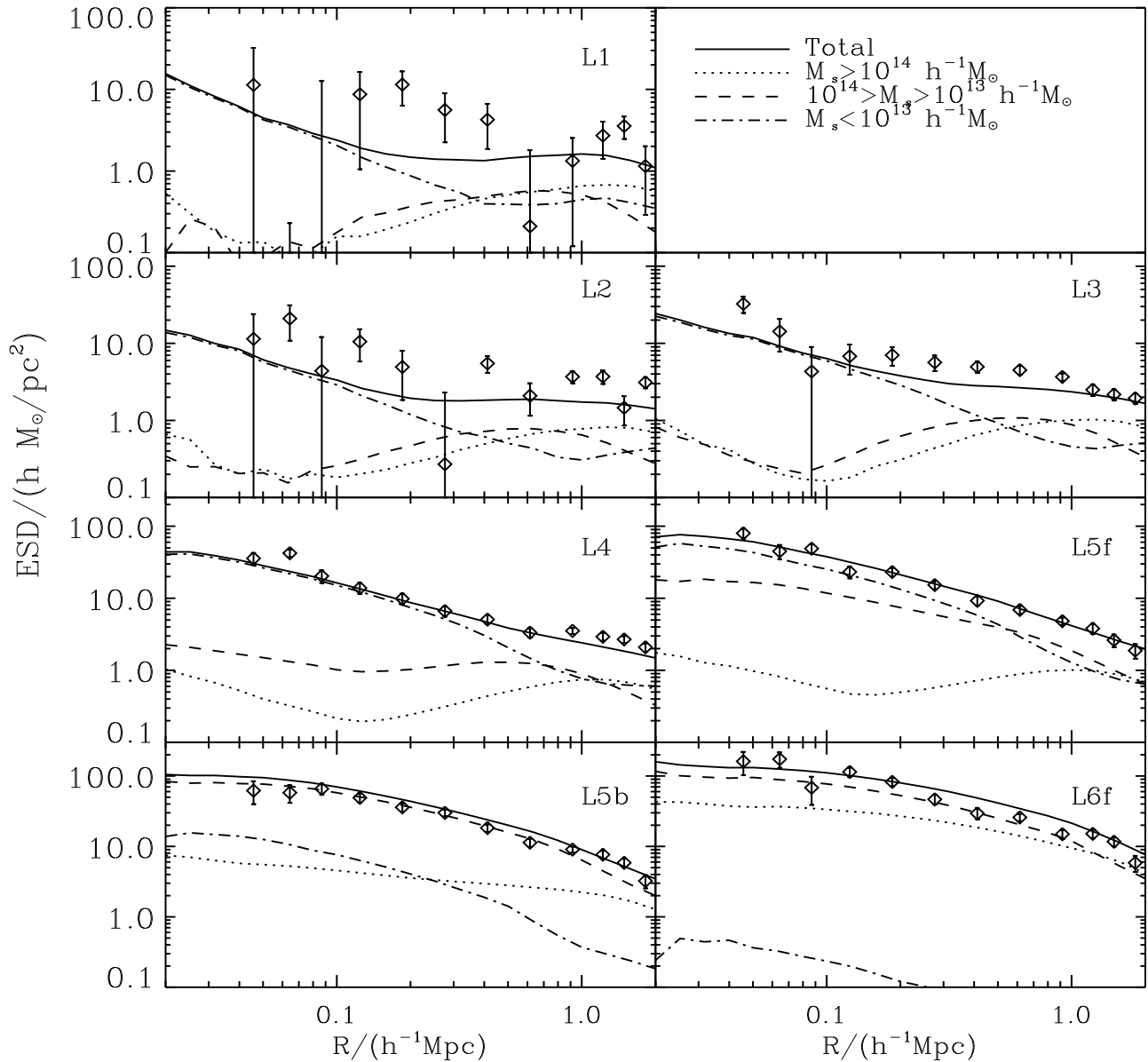}
\caption{The contribution to the ESD plotted separately for dark
  matter haloes  of different masses.  In each panel, the  dotted line
  shows       the       contribution       from      haloes       with
  $M_S\ge10^{14}h^{-1}M_{\odot}$.   The    dashed   line   shows   the
  contribution    from    haloes    with    $10^{13}h^{-1}M_{\odot}\le
  M_S<10^{14}h^{-1}M_{\odot}$,  and  the  dot-dashed  line  shows  the
  contribution  from  haloes  with $M_S<10^{13}h^{-1}M_{\odot}$.   The
  solid line shows the total  lensing signal predicted by the fiducial
  model. For  comparison, the observational data are  included as data
  points with error-bars.}\label{fig:mass}
\end{figure*}

\begin{figure*}
\includegraphics[width=\textwidth]{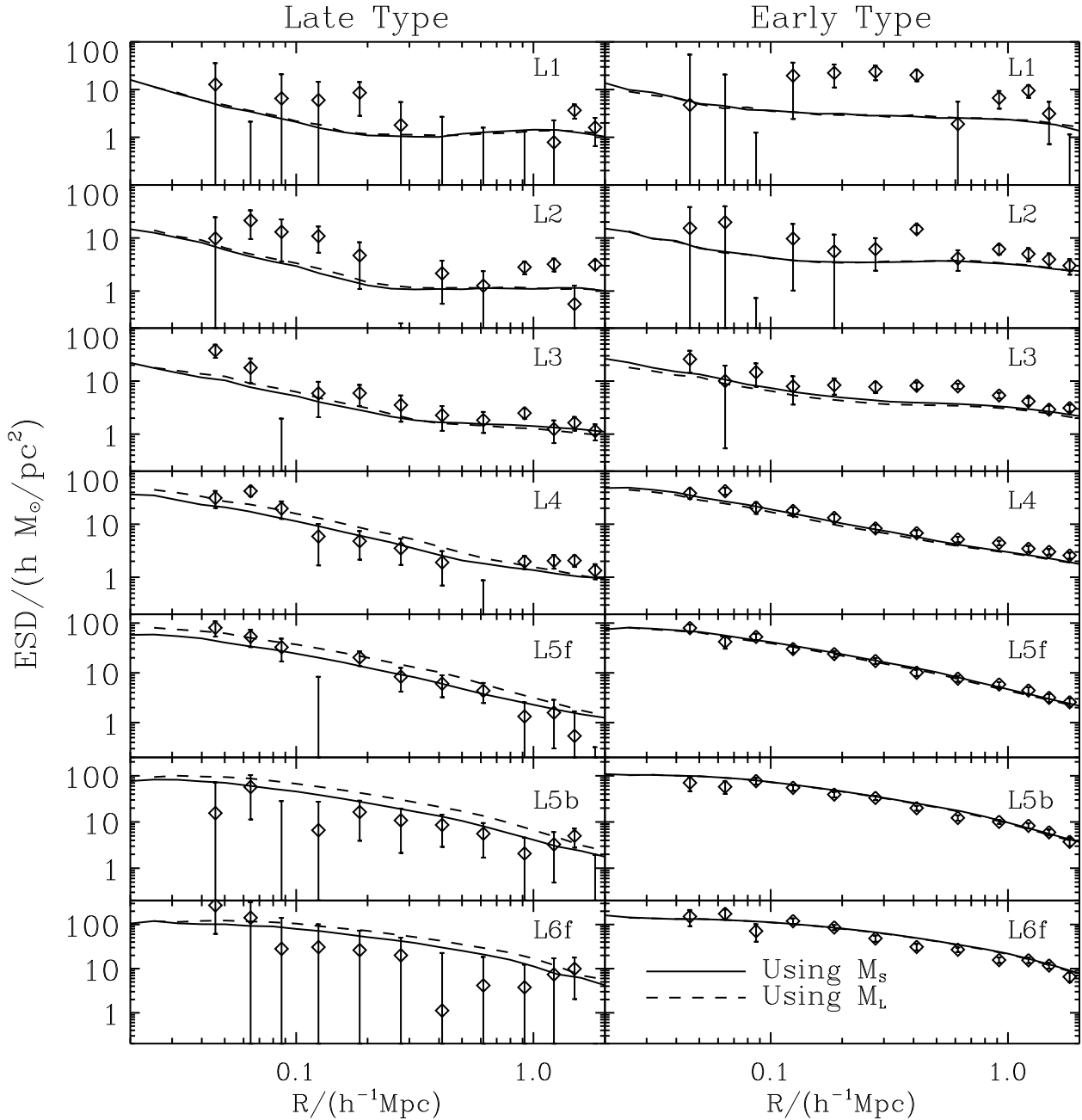}
\caption{The right panels show the ESD of early galaxies in
  different luminosity  bins, while the  left panels show  the results
  for  late  galaxies.  The  data  points  with  error-bars  show  the
  observational  results.   The model  predictions  of  the ESD  using
  stellar mass  as halo mass indicator  are shown as the  solid lines.
  For  comparison,  the  dashed  lines show  the  corresponding  model
  predictions using $M_L$ as the halo masses.  }\label{fig:4bL}
\end{figure*}

\begin{figure*}
  \includegraphics[width=\textwidth]{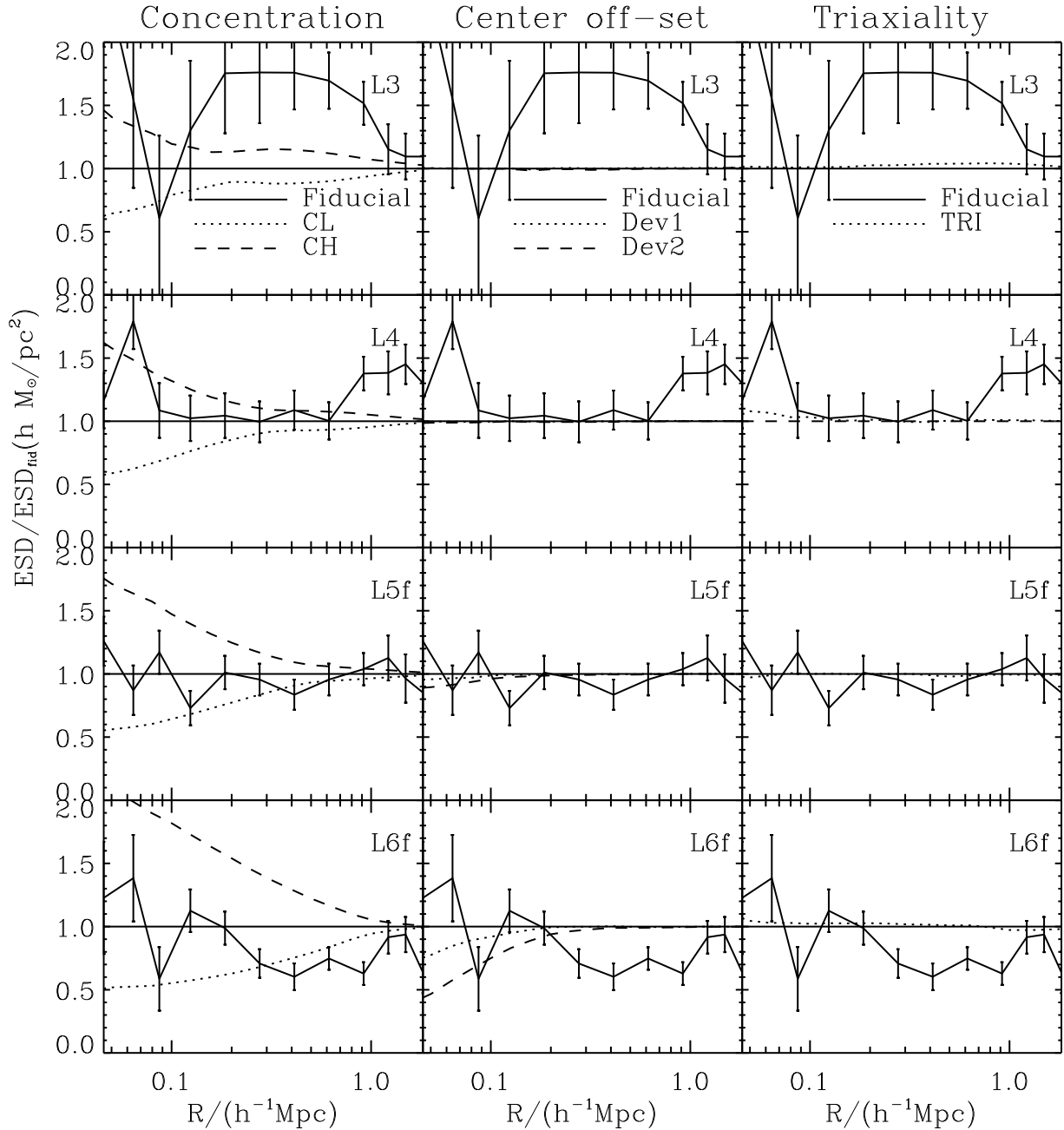}
            \caption{The
    dependence  of the  predicted $\Delta\Sigma(R)$  on  various model
    parameters. For comparison, the observational data are included as
    data points with error-bars.  The left column shows the dependence
    on halo concentration: the halo concentrations in CL (dotted line)
    and CH  (dashed line) are assumed to  be 1/2 and 2  times those in
    the fiducial  model (solid line), respectively.  The middle column
    of shows  the effects of halo  center offset. The  dotted line and
    the dashed  line show the results  of models Dev1  and Dev2 model,
    respectively,  while the solid  line is  the fiducial  model.  The
    right column  shows the effect  of assuming triaxial  halo density
    profile. The dotted line shows  the result of model TRI, while the
    solid line again shows the fiducial model. Both the observation and
    model predicted $\Delta\Sigma(R)$ are normalized by the fiducial
    prediction. }\label{fig:com}
\end{figure*}

\begin{figure*}
\includegraphics[height=0.7\textheight,width=\textwidth]{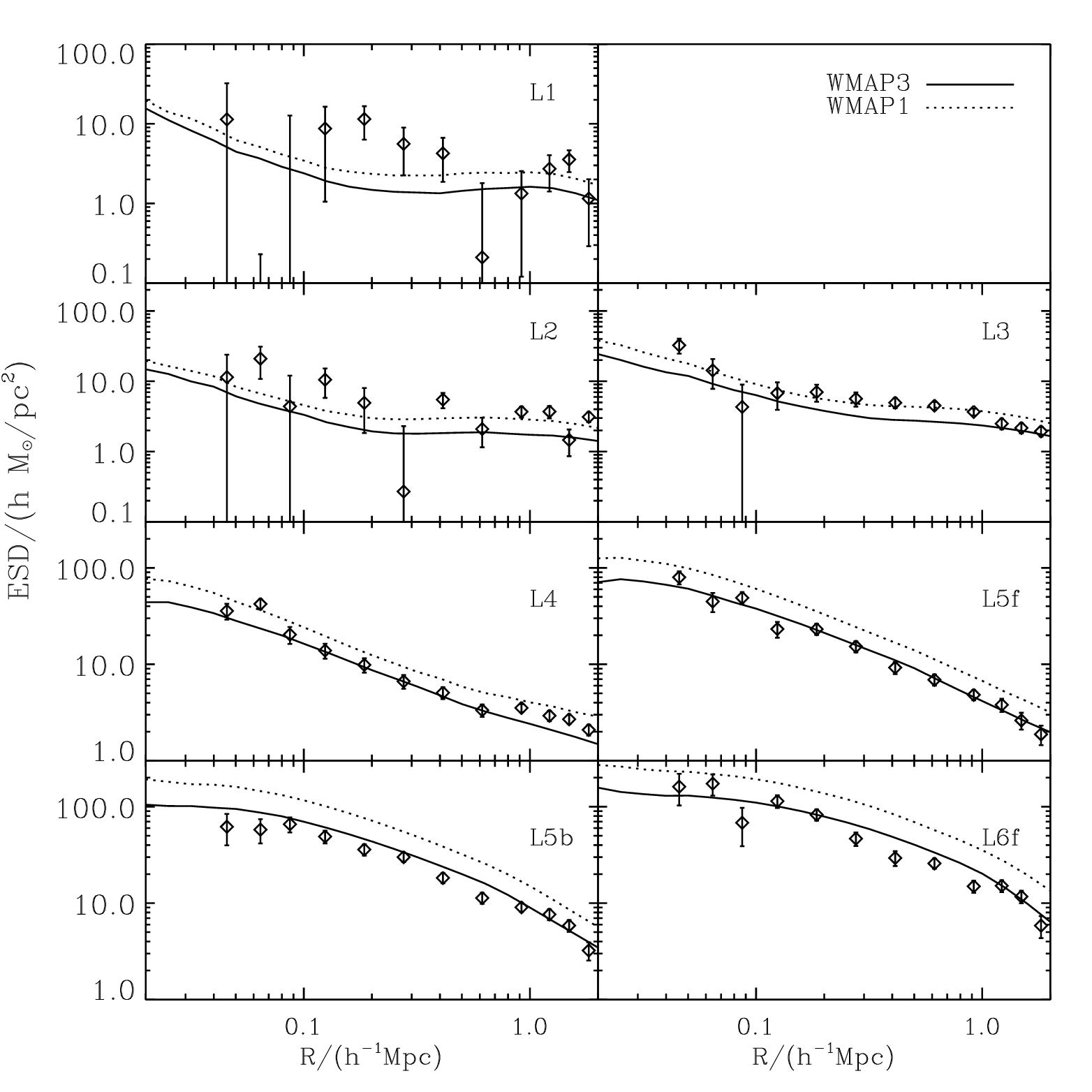} \caption{The model prediction of ESD assuming different cosmological
  models.   The solid  and dotted  lines show  the fiducial  model and
  model using WMAP1 parameters, respectively .  The observational data
  are plotted as data points with error-bars. }\label{fig:wmp}
\end{figure*}

\section{Results}
\label{sec_results}

\subsection {SDSS Lensing Data}

Before  presenting  our  model  predictions,  we  first  describe
the observational  results that  we will  use for comparison. The
observational results to  be used were obtained by Mandelbaum et al.
(2006),  who analyzed  the galaxy-galaxy lensing effects using
galaxies in  a sample constructed from  the SDSS  DR4 spectroscopic
sample. Their sample of lensing galaxies is similar to the galaxy
sample used in Y07  to construct the  group catalogue used here. The
only difference  is that Mandelbaum  et al's  sample includes  all
galaxies with redshifts in the range $0.02<z<0.35$, while the
galaxies in Y07's group catalogue are in $0.01\leq z \leq 0.2$.
Since, as to be described below, we are interested in the lensing
signals around galaxies of given luminosity and morphological type,
this difference in redshift range is not expected  to have a
significant impact on our results. For the faint luminosity bins,
our galaxy samples should be almost identical to that of Mandelbaum
et al. (2006),  because all faint galaxies are at $z \leq  0.2$ in
the SDSS catalog. For bright galaxies we also expect the statistic
properties of the two samples to be similar. Both Mandelbaum et al.
(2006) and Y07 applied similar evolution correction and K
correction, so that the evolution in the galaxy population has been
taken into account, albeit in a simple way. As we will see, even for
the the two brightest bins, the lensing signal is dominated by halos
with masses $\sim 10^{14}\,h^{-1}{\rm M}_\odot$, and the change in
the halo mass function around this mass is small between $z=0.2$ and
$0.35$. Following Mandelbaum et al. (2006), we split the galaxy
sample into 7 subsamples according to galaxy luminosity. Table
\ref{table:t1} shows  the properties of these subsamples: the
luminosity  range covered by each subsample, the mean redshift, the
mean luminosity, and the fraction of late-type galaxies. As expected
the mean redshifts of brightest bins are different from the
corresponding redshifts in Mandelbaum et al.

\begin{table}
\caption{The properties of galaxy samples. In each case, the
absolute-magnitude range,  the mean redshift,  the mean luminosity,
and the fraction of late-types are listed. Note that
$L_{*}=1.2\times 10^{10}h^{-2}L_{\odot}$}
\begin{tabular}{|c|c|c|c|c|}
\hline
 Sample & $M_r$ & $\langle z\rangle$
& $\langle L/L_{*}\rangle$ & $f_{\rm late}$\\
  \hline
L1 & $-18<M_r<-17$& 0.031& 0.075    & 0.81 \\
L2 & $-19<M_r<-18$& 0.048& 0.191    & 0.70  \\
L3 & $-20<M_r<-19$& 0.074& 0.465    & 0.54 \\
L4 &$-21<M_r<-20$ & 0.111& 1.13     & 0.35\\
L5f&$-21.5<M_r<-21$& 0.145& 2.09   & 0. 22 \\
L5b&$-21.5<M_r<-22$& 0.150& 3.22    & 0.12 \\
L6f&$-22<M_r<-22.5$& 0.152& 5.01   & 0.04 \\
\hline
\end{tabular}
\label{table:t1}
\end{table}

Also  following  Mandelbaum  et  al.  (2006), we  split  each galaxy
subsample into  two according to galaxy morphology.  The separation
is made  according to the  parameter $frac\_{\rm  dev}$ generated by
the PHOTO pipeline. The value of  $frac\_{\rm dev}$ is obtained by
fitting the  galaxy profile,  in a  given band,  to a model profile
given by $frac\_{\rm  dev}\times F_{\rm deV}+(1-frac\_{\rm
dev})\times F_{\rm
  exp}$, where $F_{\rm deV}$ and  $F_{\rm exp}$ are the de Vaucouleurs
and exponential  profiles, respectively. As  in Mandelbaum et  al.,
we use the  average of $frac\_{\rm dev}$ in  the $g$, $r$ and  $i$
bands. Galaxies  with   $frac\_{\rm  dev}  \geq  0.  5$   are
classified  as early-type, while those with $frac\_{\rm dev} < 0. 5$
as late-type.

 It should be pointed out that we did not carry a ray-tracing
simulation to predict the galaxy-galaxy lensing results. Instead, we
directly calculate the excess surface density around SDSS galaxies.
Thus, our calculation doesn't deal with source galaxies. On the
other side, in Mandelbaum et al. (2006) the source galaxies are
carefully weighted, and lensing signals are calibrated, to reduce
any bias in the observational measurements (see Mandelbaum et al.
2005 for details). Thus, we assume the observational results are
unbiased, and compare them directly with our model predictions.

 The observation data used here was kindly provided by R.
Mandelbaum. The data used in Fig. \ref{fig:4bL} has been
published in Mandelbaum et al. (2006) where lensing signal is
calculated for early and late type galaxies separately.
R. Mandelbaum also provided the lensing data combining early and late
type galaxies for us to make the comparisons presented in
all other figures. Note that here we only show the comparison
of the lensing signal for galaxies divided according to luminosity.

 The errorbars on the observational points are $1 \sigma$
statistical error. The systematic error of the galaxy-galaxy lensing
has been discussed in detail in Mandelbaum et al. (2005). The test
has been carried out for three source samples: $r<21$, $r>21$,
and high redshift LRGs. The overall systematic error is found
to be comparable to or slightly larger than the statistical error
shown here.

\subsection{The Fiducial Model}
\label{sec:FiducialModel}

In Fig.  \ref{fig:4bs}  we show the lensing signal  around galaxies
in different luminosity bins obtained  from our fiducial model,
which has model  parameters as  described in  the last  section and
assumes the WMAP3  cosmology.   Here the  ESD  is plotted  as  a
function of  the projected  distance  $R$ from  galaxies.   The
solid  line shows  the averaged ESD of all galaxies in the
corresponding luminosity bin.  The amplitude  of  the predicted  ESD
increases  with galaxy  luminosity, reflecting the fact  that
brighter galaxies on average  reside in more massive  haloes, as
shown  in Figs.~\ref{fig:ML}  and \ref{fig:dis}. These results are
to be compared with the data  points which show the observational
results  obtained by Mandelbaum  et al. (2006)  for the same
luminosity  bins. Overall,  our  fiducial  model reproduces  the
observational  data reasonably  well, especially  for  bright galaxy
bins where the observational results are  the most reliable. The
reduced $\chi^2$ is $3.2$ combining all the luminosity bins. The
best match is for L5f, with a reduced $\chi^2$ of $0.9$. Given that
we do not adjust any model parameters, the $\chi^2$ indicates a good
agreement. For the three low-luminosity bins, the predicted ESD is
lower than the corresponding observational result. For  L1 and L2,
the observational  data are very uncertain. For L3, if we take the
observational  data points at face-value, the discrepancy with the
model prediction is significant. As described in
\S~\ref{ssec_massassign}, for groups which do not contain any member
galaxies  with $M_r -5\log  h< -19.5$,  their halo masses are {\it
not} obtained from the ranking of $M_{\rm stellar}$, but from the
average  stellar  mass-halo  mass relation of  central galaxies that
is required to  match the observed stellar mass function of central
galaxies. While all the galaxies in the bright luminosity bins have
their host halo mass assigned by ranking method, the fraction of the
galaxies in halos that have masses assigned according to the
mass-halo mass relation is about $30\%$ in L3 and about $70\%$ in L2
and L1. It is possible that this relation underestimates the halo
mass. In order to see the effect caused  by such uncertainties, we
have used a set of parameters from Yang et al. (2008) that are still
allowed by the observed stellar mass function but give larger halo
masses to the hosts of faint central galaxies. This increases the
predicted ESD for L3  by $\sim 20\%$, not sufficient to explain the
discrepancy. Indeed, this discrepancy is not easy to fix. In the
observational data,  the amplitudes of the ESD for L3 at $R\sim 0.3$
-  $0.9 h^{-1}{\rm Mpc}$ are actually slightly higher than for the
brighter sample L4, while in our model the ESD for L3 is always
lower than that  for L4. There is an effect that may help to reduce
the discrepancy  between our model prediction and the observational
results for low-mass galaxies. Since the group catalogue is  only
complete down  to certain  halo mass  limit at different redshift
(see Y07), additional assumptions have to be made in order to model
the distribution of the haloes below the mass limit.   In the model
described above, we have assumed that the haloes below the mass
limit have  a random distribution, so that they do  not contribute
to the ESD.  However, in reality these low-mass haloes are
correlated with the more  massive ones. As a result, our assumption
will underestimate the 2-halo term of the ESD.

In order to understand how the predicted ESD is produced, we also
show separately the contributions from different sources.  The
dotted lines show the ESD  contributed by the 1-halo term  of
central galaxies. For all the luminosity bins, this term dominates
the ESD at small $R$. For the brightest  two bins, this term
dominates the ESD  over the entire range  of $R$  studied. This
reflects the  fact that  almost  all the galaxies  in these two bins
are  central galaxies  and the  haloes in which they  reside are
more extended. The  dashed lines show  the ESD contributed by the
1-halo term of satellite galaxies.  This term first decreases with
$R$ and then increases to a peak value before declining at large
$R$. This  owes to  the fact  that, in  the inner  part, the lensing
signal produced  by satellite  galaxies is  dominated  by the
subhaloes associated with them, while at larger $R$ the lensing
signal produced by  satellite galaxies  is dominated by  their host
halos.  The value of $R$ at which  the ESD reaches the minimum
corresponds roughly to  the  average halo-centric  distance  of the
subhaloes  in  the luminosity bin. Note that the  contribution of
satellites to the total ESD is only important at  large $R$ in the
low-luminosity bins.  This reflects the  fact that a significant
fraction of  the low-luminosity satellites reside  in massive
haloes. Note that although the 1-halo satellite term dominates the
lensing signal at $R\geq 0.3$$h^{-1} {\rm Mpc}$ for faint luminosity
bins, the discrepancy between the observation and our prediction
cannot be simply solved by boosting up the satellite contribution.
The reason is that the increase of the 1-halo satellite term
requires the increase of the host halo mass of the satellites, which
will make the 1-halo central term of the bright bins increase as
well, causing significant discrepancy for the bright bins. Finally,
the dash-dotted lines represent the contribution of the 2-halo term.
As expected, this term is relevant only on relatively large scales.
In our model this term never dominates at $R\le 2\,h^{-1}{\rm Mpc}$.
However, as discussed before this term may be underestimated here.
In C08, the 2-halo term is found to be comparable to the 1-halo
satellite term even at $R \sim 0.3  h^{-1} {\rm Mpc}$. Unfortunately
the 2-halo term in C08 may be overestimated because halo-exclusion
effect is not properly included.

The large fluctuations seen in L6f are due to the small number of
galaxies in this luminosity bin.

Since the halo  mass of each lensing galaxy is known  in our model, we
can also  examine the contributions to  the total ESD in  terms of the
halo mass. Fig. \ref{fig:mass} shows the results where the host haloes
are  split into  three  bins  of $M_S$:  $M_S\ge  10^{14} h^{-1}  {\rm
  M}_{\odot}$   (dotted  lines);  $10^{13}h^{-1}   {\rm  M}_{\odot}\le
M_S<10^{14} h^{-1}  {\rm M}_{\odot}$ (dashed  lines); and $M_S<10^{13}
h^{-1}  {\rm M}_{\odot}$ (dashed-dotted  lines). As  one can  see, the
lensing signals in brighter bins are dominated by more massive haloes.
Very massive  haloes with $M_S\ge 10^{14} h^{-1}  {\rm M}_{\odot}$ are
not the  dominant contributor, even  for the brightest  luminosity bin
considered here, because  the total number of galaxies  hosted by such
haloes are  relatively small. For the  low-luminosity bins, relatively
massive haloes  dominate the ESD  at large $R$, because  a significant
fraction of the low-luminosity satellites are hosted by massive haloes
(see Figs.~\ref{fig:ML} and \ref{fig:dis}).

\subsection {Dependence on Galaxy Type}
\label{sec:depend}

In Fig. \ref{fig:4bL} we present the results separately for
early-type and late-type galaxies.  For a given luminosity bin, the
predicted ESD has a  higher amplitude  for early-type galaxies,
clearly due  to the fact that  early-type galaxies  are more likely
to reside  in massive haloes  (see e.g.  van den  Bosch,  Yang \& Mo
2003).  For the  faint samples, L1 and  L2, the behavior of the
predicted ESD for early-type galaxies  resembles that  of  satellite
galaxies  in these  luminosity bins, while the predicted ESD for the
late-type galaxies looks like of the  central  galaxies in  the
corresponding  luminosity bins.  This, again,  reflects the fact
that faint  early-type galaxies  are mostly satellites in massive
haloes,  while the faint late-type population is dominated by the
central galaxies in low-mass haloes.

The  dashes lines  in Fig.   \ref{fig:4bL} show  the  results
obtained using    $M_L$    as   halo    mass,    rather    than
$M_S$    (see \S~\ref{ssec_massassign}). For early-type  galaxies,
the results based on this halo  mass estimate are very similar to
those  based on $M_S$. However, for the late-type samples,  the ESDs
obtained using $M_L$ are significantly higher  than those obtained
using  $M_S$, especially for the brighter samples.   This is mainly
due to  the fact that late-type galaxies  contain significant
amounts  of young  stars, so  that their stellar  mass-to-light
ratios are  relatively low.  Consequently, they are assigned a
larger halo mass based on  their luminosity than based on their
stellar mass.

The model  predictions are compared with the  observational results
of Mandelbaum et al. (2006). Here  again, the model prediction based
on $M_S$ matches the observational data for the four bright samples.
For the three faint  bins, the model predictions are again lower
than the observational results. As shown in Fig. \ref{fig:4bL}, it
seems that our model prediction agrees better with the observation
for the late type galaxies. For example, for L3 our model prediction
matches the observation reasonably well for the late-type subsample
(with the reduced $\chi^2$ equal to $1.6$), while the discrepancy is
quite large for the early-type subsample (with reduced $\chi^2$
equal to $5.3$). Since faint, early-type galaxies are preferentially
satellite galaxies in relatively massive halos while faint,
late-type galaxies are mostly central galaxies in relatively low
mass halos, the above results seem to indicate that the discrepancy
is due to the underestimate of the 1-halo satellite term in the
model. Unfortunately, there is no simple modification of the model
that can fix the discrepancy. Since the spatial distribution of
galaxies is fixed by the observation, the only change that can be
made is in halo properties. As we will see in the following
subsection, increasing the halo concentration even by a factor of
two can only increase the predicted ESD by about 20\% for the L3
sample, insufficient to explain the discrepancy. An increase in the
halo masses assigned to galaxy groups can reduce the discrepancy for
faint galaxies, but it would also significantly over-predict the ESD
for bright galaxies. The other possibility is that the halo masses
of isolated, faint early-type galaxies are significantly
underestimated. For example, isolated early-type galaxies may reside
in much more massive halos than that given by their stellar masses.
In order to explain the discrepancy, the halo masses for these
galaxies need to be larger by a factor of at least 3. This will not
affect significantly the prediction for bright galaxies, but can
boost the prediction for L3 by 50\% at $R\sim 0.2h^{-1}{\rm Mpc}$.
Unfortunately, it is still unclear if the required halo mass
increase is feasible in current models of galaxy formation.

\subsection {Dependence on other Model Parameters}
\label{sec:parameterdepend}

 The    ESD   signal    predicted    by   the    model outlined in
\S~\ref{sec:FiducialModel} is based on several assumptions. Beside
the underlying cosmology and  the halo mass assignment which are the
most crucial   model  ingredients  (see \S~\ref{sec:CosmoDependence}
and \S~\ref{sec:HaloMassAssignment}), the     model     requires a
concentration-halo mass relation and it assumes that central
galaxies reside at rest at the centre of a \textit{spherical} dark
matter halo. In  this subsection, we test how  our results  are
affected  by these assumptions.

The concentration parameter is a  measure of the amount of dark
matter in the  central regions of  the haloes. Accordingly,
different models for the  concentration-halo mass relations  are
expected to  result in different predictions for  the ESD signal (at
least  on small scales). The  fiducial model described  in
\S~\ref{sec:FiducialModel}  uses the model  of Macci\`o  et al.
(2007).  However,  other models  are also available in the
literature (e.g.  Bullock et al. 2001; Eke,  Navarro \& Steinmetz
2001), which predict concentration-mass relations that are slightly
different (see C08  for  an  assessment  of the  impact  on
galaxy-galaxy lensing).   Furthermore, the  presence  of a (central)
galaxy in a dark matter halo may have an  impact on its
concentration via, for example, adiabatic contraction (e.g.
Blumenthal et al. 1986), which is  not  accounted  for  in  the
concentration-mass  relations obtained from pure $N$-body
simulations.  Finally, attempts to measure halo  concentrations
observationally have  thus far  given conflicting results (e.g.  van
den Bosch \& Swaters 2001; Comerford \& Natarajan 2007, and
referenecs therein).  To examine how the lensing predictions depend
on changes  in halo concentation we consider  two models: model CH,
in which the concentrations are 2 times as high as in our fiducial
model, and model CL, in which  the concentrations are 2 times
smaller. The left column of Fig.~\ref{fig:com} shows the predictions
of models CH (dashed lines) and CL (dotted lines) compared to our
fiducial model (solid lines). Results  are only shown for four
luminosity bins, as indicated. Note   that  the   model   with
higher  (lower) halo concentrations  predicts ESDs  that are higher
(lower). The effect is stronger on scales where the 1-halo central
term dominates (see Fig~\ref{fig:4bs}).  Accordingly, in the case of
the brightest sample (L6f), models CL and CH  differ from the
fiducial model  on scales up to  $R\simeq 1  \, h^{-1} {\rm Mpc}$,
while in sample L4 the differences  are only appreciable out to
$R\simeq 0.5 \, h^{-1} {\rm Mpc}$. We conclude that the predicted
ESD depends quite strongly on the assumed halo concentrations,
indicating that galaxy-galaxy lensing has  the potential to
constrain the density profiles of dark matter haloes (see Mandelbaum
et al. 2008). In this paper, we have assumed that the concentration
of a halo depends only on its mass, we have ignored the possible
halo age-dependence of the concentration-mass relation (e.g.
Wechsler et al. 2002; Zhao et al. 2003; Lu et al. 2006).
If for a given mass older halos have higher concentration, and if
the formation of a galaxy in a halo depends strongly on the formation
history of the halo, then the age-dependence of the halo concentration
must be taken into account. Unfortunately, it is unclear how to
connect the halo age (hence halo concentration) with the properties of
galaxies. For a given halo mass, the dispersion in the concentration
is about 0.12 dex (e.g. Jing 2000), which corresponds to a change of
about 20\% in the predicted ESD on  $R\sim 0.1  \, h^{-1} {\rm Mpc}$.

In our fiducial  model, central galaxies are assumed  to reside at the
center  of their dark  matter haloes.   However, as  shown in  van den
Bosch  et al.   (2005), in  haloes with  masses $M_h>10^{13}h^{-1}{\rm
  M}_{\odot}$ there  is evidence to suggest that  central galaxies are
offset  from their  halo centers  by $\sim  3$ percent  of  the
virial radius. A similar  result was obtained by Berlind  et al.
(2003) using SPH simulations  of galaxy  formation. To examine  how
such  an offset impacts on the ESDs, we  consider two additional
models. Following van den  Bosch et  al. (2005)  and  Yang et  al.
(2006),  we assume  that central galaxies  in haloes with $M_S  >
10^{13}h^{-1}{\rm M}_{\odot}$ are offset from  their halo centers by
an amount that  is drawn from a Gaussian distribution with zero
mean.  We use two different values for the dispersion of the
distribution:  $3\%$ of the virial radius (model `Dev1')  and $6\%$
of  the  virial  radius   (model  `Dev2').   The corresponding
lensing  predictions are shown  in the middle  column of
Fig.~\ref{fig:com} as the dotted (Dev1) and dashed (Dev2) lines.
Note that  the offsets  only affect  the lensing  signal for  the
brightest sample (L6f), where a larger  offset results in a stronger
suppression of  the ESD  on small  scales ($R  < 0.1\,  h^{-1}{\rm
Mpc}$). For fainter samples,  no (significant)  differences with
respect to the fiducial model  are apparent,  which owes  to  the
fact  that fainter  centrals typically reside  in haloes with $M_h <
10^{13}h^{-1}{\rm M}_{\odot}$ which do not have an offset (at least
in our models).

As  a  final  test, we  examine  the  impact  of  halo shapes  on
the galaxy-galaxy  lensing signal.   In  our fiducial  model, dark
matter haloes  are assumed  to be  spherically symmetric.   However,
$N$-body simulations  show that,  in  general, they  are  triaxial
rather  than spherical.   Jing  \&  Suto  (1998)  proposed a fitting
formula  for triaxial dark matter haloes, which has been applied to
both strong and weak lensing  analyses (e.g.   Oguri, Lee \& Suto
2003; Oguri \& Keeton 2004; Tang  \& Fan 2005).   In order  to
examine  the impact of our assumption  of halo sphericity, we
consider an alternative model (model  `TRI'), in which we  assume
that  the dark matter density distribution  is given  by $\rho_{\rm
TRI}(R)$, where $R$ specifies an ellipsoidal surface:
\begin{equation}
R^2=\left(\frac{x^2}{a^2}+\frac{y^2}{b^2}+\frac{z^2}{c^2}\right)c^2\,.
\end{equation}
Here $a\le b\le c$ are the three principal semi-axes of the
ellipsoid. We set $\rho_{\rm TRI}(R)= \rho (R')$, where $\rho(R')$
is the NFW profile, so that the total mass within a sphere of radius
$R'$ in $\rho(R')$ is equal to the mass within the elliptical shell
at $R$. For the axis ratios we adopt the distribution function given
by Jing \& Suto (2002):
\begin{eqnarray}\label{pac}
  p(a/c)& = &\frac{1}{\sqrt{2 \pi}\times 0.113}
\left(\frac{M_{\rm vir}}{M_{NL}}\right)^{0.07[\Omega(z)]^{0.7}}\nonumber \\
    & \times & \exp\left\{-\frac{\left[(a/c)(M_{\rm
        vir}/M_{NL})^{0.07[\Omega(z)]^{0.7}} -0.54\right]^2}
    {2(0.113)^2}\right\} \nonumber \\
\end{eqnarray}
and
\begin{eqnarray}\label{pab}
  p(a/b|a/c)&=&\frac{3}{2(1-\max(a/c,0.5))}\nonumber \\
  & \times &\left[1-\left(\frac{2a/b-1-\max(a/c,0.5)}{1-\max(a/b,0.5)}
    \right)^2\right]\,,
\end{eqnarray}
where $M_{NL}$ is the characteristic mass  scale, on which the rms
of the top-hat  smoothed over-density is  equal to  $1.68$.  In
practice, we proceed as follows:  For each dark matter halo we first
draw the axis ratio   $a/c$  and   $a/b$  using Eqs.~(\ref{pac})
and~(\ref{pab}), respectively. Next  we draw a random 3D orientation
of  the principal axes,  and project  the dark matter  particles
along the  (fiducial) line-of-sight.  Next, for each halo,  we
determine the major  axis of the projected distribution  which we
align with the  major axis of the central galaxy.  This assumption
is motivated by observational claims that the  major axis of a
central  galaxy tends to aligned with  that of its host halo (e.g.
Yang et al.  2006; Faltenbacher et al.  2008; Wang et al.  2008). As
shown in  the right-hand panels of Fig. \ref{fig:com}, changing from
spherical  haloes  (solid lines)  to triaxial  haloes (dotted lines)
has  almost no  impact  on the predicted ESDs.   This should not
come as an  entire surprise, since the ESDs are azimuthally averaged
over many haloes, which have random orientations on the sky. Note that
here we assume that the halo is perfectly aligned with the central galaxy.
However, the observational results mentioned above actually suggest a
misalignment. We have tried a model in which the orientation of the host
halo is uncorrelated with that of the central galaxy. The change in the
results is very small compared with the model assuming perfect
alignment.

\subsection{Dependence on Cosmology}
\label{sec:CosmoDependence}

Another  very important  assumption  in our  model  prediction is  the
cosmological model in  the calculations of the halo  mass function and
the  geometrical properties  of spacetime.  The redshifts  of galaxies
considered  here are  restricted  to  $z\le 0.2$,  and  the impact  of
changing cosmological  parameters on  the spacetime geometry  is quite
small  in  our analysis.  On  the  other  hand, changing  cosmological
parameters can  change the halo masses assigned  to individual groups,
which may have  significant impact on the expected  lensing signal. In
our fiducial  model, we adopt  the cosmology parameters  obtained from
the    WMAP    3-year   data,    with    $\Omega_{\rm   m}=0.    238$,
$\Omega_{\Lambda}=0.762$,  $n=0.951$, and $\sigma_8=0.75$  (Spergel et
al. 2007).  As comparison, we will show some results obtained assuming
another  set of  cosmological parameters  with $\Omega_{\rm  m}=0. 3$,
$\Omega_{\Lambda}  = 0.7$,  $n =  1.0$, and  $\sigma_8=0.9$,  which is
strongly supported by the first  year data release of the WMAP mission
(see Spergel  et al.  2003) and has  been considered in  many previous
studies.  In  what  follows  we  will  refer to  this  second  set  of
parameters  as  the  WMAP1   cosmology.  Note  that  the  cosmological
parameters given by the recent WMAP 5-year data (Komatsu et al.  2008)
are in between those of WMAP1 and WMAP3.

Fig. \ref{fig:wmp}  compares the ESD  predicted by the  fiducial
model using  the WMAP3  cosmology (solid  lines) and  that predicted
by the WMAP1 cosmology (dotted  lines). As one can see,  the ESD
predicted by WMAP1 is significantly higher than that predicted by
WMAP3, especially for bright galaxies.   Most of this increase is
due  to changes in the halo mass  function, which  causes (massive)
groups  to be  assigned a larger  halo mass.   The changes  in the
halo concentrations  and the spacetime geometry play only a  minor
role. A comparison with the SDSS data  clearly favors  the WMAP3
cosmology over  the  WMAP1 cosmology, especially for  the brighter
luminosity  bins. The reduced $\chi^2$ for the WMAP1
cosmology is $21.3$, much larger than $3.2$ for the WMAP3 cosmology.
This result is in good agreement with that obtained in C08 using the
CLF model.

Thus, we  conclude that the  galaxy-galaxy lensing data  either prefer
the WMAP3  cosmology, which  has a relatively  low $\sigma_8$,  or our
halo mass assignment is in serious error.  In the following subsection
we  show that  the  uncertainties  in our  halo  mass assignments  are
unlikely to  change our results significantly.   We therefore conclude
that the galaxy-galaxy lensing data prefer a $\Lambda$CDM model with a
relatively  low $\sigma_8$.   If we  use WMAP5  parameters,  the model
prediction is in  between WMAP1 and WMAP3, which is  still too high to
matched the observed ESD of bright galaxies.

\subsection{Uncertainties in Halo-Mass Assignment}
\label{sec:HaloMassAssignment}

In  our model,  the masses  of groups  are assigned  according  to
the stellar-mass  ranking and  the  halo mass  function  predicted
by  the adopted cosmology. The underlying assumption is that the
mass function of the  host haloes  of groups is  the same  as that
predicted  by the cosmological model. However, even if the
cosmological model adopted is a good approximation  to the real
universe, the  observed halo density may be different from the model
prediction because of cosmic variance introduced  by the finite
observational volume.   The effect  of such variance  is expected to
be  most  important for  massive haloes, because the number density
of such systems is small.  If, for example, the number density  of
massive haloes in the  observational sample is, due to cosmic
variance, smaller  than the model prediction,  the mass assignment
with  the use of  the theoretical halo mass function would assign a
higher  halo mass to groups. Consequently, the ESD of bright
galaxies,   which  are   biased  toward   massive haloes, would  be
overestimated.  Here  we  test  the  importance of  such effects  by
considering the uncertainties due  to Poisson fluctuations. We use
the halo mass  function predicted with  the WMAP3 cosmology to
generate a set of random halo samples, each  of which contains the
same number of groups as the  observational sample. The halo masses
in each of these samples is  then ranked in descending  order and
the  halo mass  of a given rank is then assigned to the group with
the same rank in stellar mass.  We  find that the  scatter in the
ESD obtained in this  way is negligibly small, even  for the
brightest sample. The  reason is that, even for the brightest
sample, the ESD  is dominated by  haloes with intermediate   masses,
$10^{13}h^{-1}{\rm   M}_\odot   <    M_S < 10^{14}h^{-1}{\rm
M}_\odot$ (see Fig.\, \ref{fig:mass}), and the total number of such
groups is quite large.

Another  uncertainty  in the  mass  assignment  may  arise from
fiber collisions.  In the SDSS survey, no  two fibers on the same
SDSS plate can  be  closer  than   55  arcsec.   Although  this
fiber  collision constraint is partially alleviated by the fact that
neighboring plates have overlap  regions, $\sim 7$  percent of all
galaxies  eligible for spectroscopy do  not have a  measured
redshift.  Our analysis  here is based  on the  group catalog
constructed from  galaxy Sample  II (see Y07), in which many of the
galaxies missed due to fiber collisions are not  included.
Consequently,  the total  stellar mass  (or  the total luminosity)
of  some  of  the  groups  may  be  underestimated,  thus
introducing  a bias  in  the ranking.   This  bias is  expected to
be stronger in  richer systems because they have  higher projected
galaxy number density  and are more likely  to suffer from  fiber
collisions. However,  this effect  is  not likely  to  have a  big
impact on  our results, because  it only changes  the relative
ranking of  the groups that  have   similar  stellar   mass
(luminosity)  and   because  the galaxy-galaxy lensing signals are
averaged in relatively broad bins of galaxy luminosity.

In order to quantify this effect we carry out a similar analysis using
the group catalog constructed from Sample III, where a galaxy affected
by fiber collisions  is assigned the redshift of  its nearest neighbor
(see Y07 for details). In this  case, the situation is the opposite to
that in Sample II, because  here some galaxies may be wrongly assigned
to  groups  in the  foreground  or the  background  due  to the  wrong
redshifts  assigned  to some  of  the  fiber-collision galaxies.   The
stellar mass and  the luminosity of some of  the groups will therefore
be  overestimated. In our  test, we  use Sample  III to  construct the
group and to  set mass to groups according to  their ranking in Sample
III, but we calculate the ESD  only around galaxies that are in Sample
II. The results  obtained in this way are very  similar to those based
on   Sample  II  alone,   suggesting  that   the  effect   of  missing
fiber-collision  galaxies in  the mass  assignment is  not  important.
However, if  the fiber-collision  galaxies are themselves  included in
the calculation of the ESD, i.e.   if we use galaxies in Sample III to
calculate the  ESD, the amplitude  of the ESD is  significantly larger
than that  obtained from  the galaxies in  Sample II.  The  reason for
this  is   that  the  galaxies   affected  by  fiber   collisions  are
preferentially located in high density  regions, so that they are more
likely associated  with a massive  halo. The observational  results of
Mandelbaum   et  al.   (2006)   are  based   on  galaxies   that  have
spectroscopic  redshifts,  and  so   a  fair  comparison  between  the
observational results and our model  predictions can only be made with
the  use  of  galaxies  in   Sample  II.   Our  test  above  therefore
demonstrates  that our  conclusions about  the comparison  between the
observational data  and the model  predictions are robust  against the
uncertainties due to fiber collisions.

  The scatter in the relation between halo mass and stellar mass
(or luminosity) may also produce some uncertainties in our model
prediction. As shown in Mandelbaum \& Seljak (2007), the scatter is
in partial degeneracy with cosmology model. In our investigation, we
have fixed cosmological parameters and allowed no dispersion in the
halo mass-total stellar mass relation. If, for example, we assume a
log normal distribution with a dispersion of 0.3 index in halo mass
for a given stellar mass, the predicted ESD would be a few percent
larger than that predicted by the fiducial model.

\section{Summary}
\label{sec_conclusions}

In this paper  we model the galaxy-galaxy lensing  signal expected
for SDSS  galaxies, using  the galaxy  groups  selected from  the
SDSS  to represent the dark matter haloes  within which the galaxies
reside. We use the properties of the dark halo population, such as
mass function, density profiles  and shapes,  expected from the
current $\Lambda$CDM cosmogony to model the dark  matter
distribution in each of the groups identified  in the  SDSS volume.
The use  of the  real  galaxy groups allows us to predict  the
galaxy-galaxy lensing signals separately for galaxies  of  different
luminosity,  morphological  types,  and  in different  environments
(e.g. central  versus satellite  galaxies). We check  the robustness
of  our  model  predictions  by  changing  the assumptions about the
dark  matter distribution in  individual groups (such as the shape,
density  profile, and center offset of dark matter haloes), as  well
as  the  cosmological  model  used to predict  the properties of the
halo population.  We compare  our model predictions with the
observational data of  Mandelbaum et al. (2006)  for similar samples
of lens galaxies. Although there is some discrepancy for lens galaxies
in the low-luminosity bins, the overall observational results can be well
understood in the current $\Lambda$CDM cosmogony.
In particular, the observed
results can be well reproduced in a $\Lambda$CDM model with
parameters based on the WMAP3 data, but a $\Lambda$CDM model with a
significantly higher $\sigma_8$, such as the one  based on the WMAP1
data, significantly over-predicts the galaxy-galaxy lensing signal.
Our results  also suggest that, once a correct model of structure
formation is adopted,  the halo masses assigned to galaxy groups
based on ranking their  stellar masses with the  halo mass function,
are statistically  reliable. The results obtained imply that
galaxy-galaxy  lensing  is  a powerful tool  to constrain both the
mass  distribution associated with galaxies  and cosmological
models. In the  future, when deep imaging surveys provide more
sources  with high  image quality in the SDSS sky coverage, the
galaxy-galaxy  lensing signals produced by the SDSS galaxies  can be
estimated to much higher accuracy. The same analysis as presented
here is expected to provide stringent constraints on the properties
of the dark  matter haloes associated with galaxies  and galaxy
systems, as well as on cosmological parameters.

\section*{Acknowledgments}

We  thank Rachel  Mandelbaum for  providing the  SDSS lensing  data
in electronic format. Part of the  computation was carried out on
the SGI Altix 330  system at the  Department of Astronomy, Peking
University. Li Ran is supported by the National Scholarship from
China Scholarship Council. XY is  supported by the {\it One  Hundred
Talents} project of the Chinese Academy  of Sciences and grants from
NSFC (Nos. 10533030, 10673023).   HJM  would  like   to  acknowledge
the  support  of  NSF AST-0607535, NASA  AISR-126270 and NSF
IIS-0611948.   Zuhui Fan would like  to acknowledge the  supports
from  NSFC under  grants. 10373001, 10533010, and 10773001, and 973
Program (No.  2007CB815401).



\begin{thebibliography}{}

\bibitem[\protect\citeauthoryear{Adelman-McCarthy}{2006}]{Ade06}
Adelman-McCarthy J. K., et al., 2006, ApJS, 162, 38
\bibitem[\protect\citeauthoryear{Bell}{2003}]{Bell03}
Bell E. F. , McIntosh D. H., Katz N., Weinberg M. D., 2003, ApJS,
149, 289
\bibitem[\protect\citeauthoryear{Berlind}{2002}]{BW02}
Berlind, A. A.,  Weinberg D. H., 2002, ApJ, 575, 587
\bibitem[\protect\citeauthoryear{Berlind}{2003}]{}
Berlind A. A., et al., 2003, ApJ, 593, 1
\bibitem[\protect\citeauthoryear{Blanton}{2003}]{}
Blanton M. R., et al. , 2003, AJ, 125, 2348
\bibitem[\protect\citeauthoryear{Blanton}{2005}]{}
Blanton M. R., et al., 2005, AJ, 129, 2562
\bibitem[\protect\citeauthoryear{Blumenthal}{1986}]{}
Blumenthal G. R., Faber S. M., Flores R., Primack J. R., 1986, ApJ,
301, 27
\bibitem[\protect\citeauthoryear{Brainerd}{1996}]{Br}
Brainerd T. G,  Blandford R. D., Smail I., 1996,  ApJ, 466, 623
\bibitem[\protect\citeauthoryear{Bullock}{2001}]{Bullock01}
Bullock J. S., Kolatt T. S., Sigad Y., Somerville R. S., Kravtsov A. V., Klypin A. A., Primack J. R., Dekel A., 2001 \mnras, 321, 559
\bibitem[\protect\citeauthoryear{Cacciato}{2008}]{}
Cacciato M., van den Bosch F. C., More S., Li R., Mo H. J, Yang X.,
2008, submitted to MNRAS (C08)
\bibitem[\protect\citeauthoryear{cole}{2000}]{Cole00}
Cole S., Lacey C.G., Baugh C.M., Frenk C.S. , 2000, MNRAS, 319, 168
\bibitem[\protect\citeauthoryear{coll}{2001}]{Coll01}
Colless M., et al., 2001, MNRAS, 328, 1039
\bibitem[\protect\citeauthoryear{Comerford}{2007}]{}
Comerford J. M., Natarajan P., 2007, \mnras, 379, 190
\bibitem[\protect\citeauthoryear{Cooray}{2002}]{}
Cooray A.,  Sheth R.,  2002,  PhR,  372,  1\
\bibitem[\protect\citeauthoryear{Cooray}{2006}]{C06}
Cooray A., 2006, MNRAS, 365, 842
\bibitem[\protect\citeauthoryear{Croton}{2006}]{}
Croton D. J., et al., 2006, \mnras, 367, 864
\bibitem[\protect\citeauthoryear{deVau}{deVau}]{}
 de Vaucouleurs G., de Vaucouleurs A., Corwin H.G., Buta R.J., Paturel G., Fouque P., 1991,
Third Reference Catalogue of Bright Galaxies, (Springel-Verlag
Heidelber)
\bibitem[\protect\citeauthoryear{Eke}{2001}]{ENS01}
Eke V.R., Navarro J.F., Steinmetz M., 2001, \apj, 554, 114
\bibitem[\protect\citeauthoryear{Faltenbacher}{2007}]{}
Faltenbacher  A. , Li  C., Mao  S., van den Bosch  F. C., Yang  X.,
Jing  Y.  P., Pasquali A., Mo H. J. , 2007, ApJ, 662, 71
\bibitem[\protect\citeauthoryear{Giocoli}{2008}]{}
Giocoli C., Tormen G., van den Bosch F.C.,  2008, MNRAS, 386, 2135
\bibitem[\protect\citeauthoryear{Gao}{2004}]{}
Gao L.,  White S. D. M.,  Jenkins A.,  Stoehr F.,  Springel V.,
2004,  \mnras,  355,  819
\bibitem[\protect\citeauthoryear{Hayashi}{2003}]{}
Hayashi E.,  Navarro J. F.,  Taylor J. E.,  Stadel J.,  Quinn T.,
2003,  ApJ,  584,  541
\bibitem[\protect\citeauthoryear{Henry}{2000}]{H00}
Henry, J. P., 2000, ApJ, 534, 565

\bibitem[\protect\citeauthoryear{Hoekstra}{2003}]{}
Hoekstra H., Franx M., Kuijken K., Carlberg R. G., Yee H. K. C.,
2003, \mnras, 340, 609

\bibitem[\protect\citeauthoryear{Hoekstra}{2004}]{}
Hoekstra  H.,  2004,  MNRAS 347,  1337

\bibitem[\protect\citeauthoryear{Hudson}{1998}]{}
Hudson M. J., Gwyn S. D. J., Dahle H., Kaiser N., 1998, ApJ, 503,
531

\bibitem[\protect\citeauthoryear{Jing}{1998}]{JS98}
Jing Y. P., Suto, Y., 1998, ApJ, 494L, 5

\bibitem[\protect\citeauthoryear{Jing}{00}]{JS00}
Jing Y. P., 2000, ApJ, 535, 30

\bibitem[\protect\citeauthoryear{Jing}{2002}]{JS02}
Jing Y. P., Suto Y.,  2002, ApJ, 574, 538

\bibitem[\protect\citeauthoryear{JMB}{1998}]{JMB98}
Jing Y. P.,  Mo H. J.,  B\"{o}rner G.,  1998,  \apj ,  494,  1

\bibitem[\protect\citeauthoryear{Johnston}{2007}]{}
Johnston D. E.,  et al., 2007, preprint (arXiv: astro-ph/0709.1159)

\bibitem[\protect\citeauthoryear{Kang}{2005}]{}
Kang X., Jing Y. P., Mo H. J., B\"{o}rner G., 2005, \apj, 631, 21

\bibitem[\protect\citeauthoryear{}{}]{}
Kauffmann G.,  White S. D. M.,  Guiderdoni B.,  1993,  MNRAS, 264,
201

\bibitem[\protect\citeauthoryear{}{}]{}
Kauffmann G. ,  White S. D. M. ,  Heckman T. M. ,  Menard B. ,
 Brinchmann J. ,  Charlot S. ,  Tremonti C. ,  Brinkmann J.  2004,
 MNRAS,  353,  713

\bibitem[\protect\citeauthoryear{}{}]{}
Katz N.,  Weinberg D. H.,  Hernquist L.,  1996,  ApJS,  105,  19

\bibitem[\protect\citeauthoryear{}{}]{}
Komatsu E. , Dunkley J. , 2008, preprint (arXiv: astro-ph/0803.0547)

\bibitem[\protect\citeauthoryear{Lin}{2007}]{Lin07}
Limousin M, Kneib J. P., Bardeau S., Natarajan P., Czoske O., Smail
I., Ebeling H., Smith G.P., 2007, A\&A, 461, 881

\bibitem[\protect\citeauthoryear{Lu}{2006}]{Lu06}
Lu Y., Mo H. J., Katz N., Weinberg M. D., 2006 , \mnras, 368 1931

\bibitem[\protect\citeauthoryear{}{}]{}
Macci\`o A. V., Dutton A. A., van den Bosch F. C., Moore B., Potter
D., Stadel J., 2007, \mnras, 378, 55.

\bibitem[\protect\citeauthoryear{}{}]{M05}
Mandelbaum R., et al., 2005, MNRAS, 361, 1287

\bibitem[\protect\citeauthoryear{}{}]{M06}
Mandelbaum R., Seljak U.,  Kauffmann G.,  Hirata C. M., Brinkmann
J., 2006 MNRAS, 368, 715

\bibitem[\protect\citeauthoryear{}{}]{}
Mandelbaum R., Seljak, U., 2007, JACP, 06, 024

\bibitem[\protect\citeauthoryear{}{}]{}
Mandelbaum R., Seljak, U., Hirata C. M.,  2008, preprint (arXiv:
astro-ph/0805.2552)

\bibitem[\protect\citeauthoryear{}{}]{}
McKay T. A.,  Sheldon  E. S.,  Racusin  J.,  et al.  2001, preprint
(arXiv: astro-ph/0108013)

\bibitem[\protect\citeauthoryear{}{}]{}
McKay T. A.,  et al.,  2002,  ApJ,  571,  L85

\bibitem[\protect\citeauthoryear{}{}]{NS97}
Nakamura T. T., Suto Y., 1997, PThPh, 97, 49

\bibitem[\protect\citeauthoryear{Natarajan}{2002}]{Natarajan}
Natarajan P., Kneib J.P., Smail I., 2002, \apj, 580, L11

\bibitem[\protect\citeauthoryear{}{}]{NFW97}
Navarro J.F., Frenk C.S., White S.D.M., 1997, ApJ, 490, 493

\bibitem[\protect\citeauthoryear{}{}]{}
Oguri  M.,  Lee  J., Suto Y., 2003, ApJ, 599, 7

\bibitem[\protect\citeauthoryear{}{}]{}
Oguri M., Keeton  C., 2004, ApJ, 610, 663

\bibitem[\protect\citeauthoryear{}{}]{PS00}
Peacock J. A., Smith R. E., 2000, MNRAS, 318, 1144

\bibitem[\protect\citeauthoryear{}{}]{}
Pearce F. R.,  Thomas P. A.,  Couchman H. M. P.,  Edge A. C., 2000,
MNRAS,  317,  1029

\bibitem[\protect\citeauthoryear{}{}]{}
Petrosian V., 1976, ApJ, 209, L1
\bibitem[\protect\citeauthoryear{}{}]{}
Saunders W., et al., 2000, MNRAS, 317, 55
\bibitem[\protect\citeauthoryear{}{}]{}
Scranton R., 2003, MNRAS 339, 410

\bibitem[\protect\citeauthoryear{}{}]{}
Schlegel D.J., Finkbeiner D.P., Davis M., 1998, ApJ, 500, 525

\bibitem[\protect\citeauthoryear{}{}]{}
Sheldon E. S., et al., 2004, AJ, 127, 2544

\bibitem[\protect\citeauthoryear{}{}]{}
Sheldon E. S., et al., 2007a, preprint (arXiv:astro-ph/0709.1153)

\bibitem[\protect\citeauthoryear{}{}]{}
Sheldon E. S., et al., 2007b, preprint (arXiv: astro-ph/0709.1162)

\bibitem[\protect\citeauthoryear{}{}]{}
Somerville R. S.,  Primack J. R.,  1999,  MNRAS,  310,  1087

\bibitem[\protect\citeauthoryear{}{}]{SP03}
Spergel D.  N., et al.,  2003, ApJS, 148,  175

\bibitem[\protect\citeauthoryear{}{}]{WMAP3}
Spergel D. N., et.al., 2007, ApJS, 170, 377

\bibitem[\protect\citeauthoryear{}{}]{}
Springel V.,  2005,  MNRAS,  364,  1105

\bibitem[\protect\citeauthoryear{}{}]{}
Springel V. ,  et al. ,  2005,  Nature,  435,  629

\bibitem[\protect\citeauthoryear{}{}]{}
Strauss M.A., et al., 2002, AJ, 124, 1810

\bibitem[\protect\citeauthoryear{}{}]{}
Tang J. Y. , Fan Z. H., 2005, \apj, 635, 60

\bibitem[\protect\citeauthoryear{}{}]{}
Tinker J.L., Weinberg D.H., Zheng Z., Zehavi I., 2005, ApJ, 631, 41

\bibitem[\protect\citeauthoryear{}{}]{}
Tyson J. A.,  Valdes F.,  Jarvis J. F.,  Mills A. P.  Jr.,  1984,
ApJ,  281L, 59

\bibitem[\protect\citeauthoryear{}{}]{vo06}
Vale A., Ostriker J. P., 2006, MNRAS, 371, 1173

\bibitem[\protect\citeauthoryear{}{}]{B02}
van den Bosch F. C. , Swaters R. A., 2001, MNRAS, 325, 1017

\bibitem[\protect\citeauthoryear{}{}]{B02}
van den Bosch F. C., 2002, MNRAS, 332, 456

\bibitem[\protect\citeauthoryear{}{}]{BYM03}
van den Bosch F. C.,  Yang X. ,  Mo H. J. ,  2003,  \mnras ,  340,
771

\bibitem[\protect\citeauthoryear{}{}]{}
van den Bosch F. C.,  Norberg P.,  Mo H. J.,  Yang X.,  2004, MNRAS,
352,  1302

\bibitem[\protect\citeauthoryear{}{}]{}
van den Bosch F. C.,  Weinmann S. M., Yang X., Mo H. J., Li C., Jing
Y.P.,  2005, MNRAS,  361,  1203

\bibitem[\protect\citeauthoryear{}{}]{B05}
van den Bosch F. C.,  Tormen G.,  Giocoli C.,  2005,  MNRAS, 359,
1029

\bibitem[\protect\citeauthoryear{}{}]{}
van den Bosch F. C., et al.,  2007, MNRAS, 376, 841

\bibitem[\protect\citeauthoryear{}{}]{}
Wang Y., Yang X., Mo H. J., Li C., van den Bosch F. C., Fan Z. H.,
Chen X., 2008, \mnras, 385, 1511

\bibitem[\protect\citeauthoryear{}{}]{}
Warren M. S., Abazajian K., Holz D. E., Teodoro L., 2006, ApJ, 646,
881
\bibitem[\protect\citeauthoryear{}{}]{}
Wechsler R. H., Bullock J. S., Primack J. R., Kravtsov A. V., Dekel
A., 2002, ApJ, 568, 52
\bibitem[\protect\citeauthoryear{}{}]{} Weinberg D. H.,
Colombi S., Dav\`e R., Katz N., 2008, ApJ, 678, 6

\bibitem[\protect\citeauthoryear{}{}]{}
White S. D. M.,  Frenk C.,  1991,  ApJ,  379,  52

\bibitem[\protect\citeauthoryear{}{}]{YMW03}
Yan R., Madgwick D. S., White M. , ApJ, 2003, 598, 848

\bibitem[\protect\citeauthoryear{}{}]{YMB03}
Yang X.,  Mo  H. J.,  van  den  Bosch F. C.,  2003,  \mnras , 339,
1057

\bibitem[\protect\citeauthoryear{}{}]{Y05}
Yang X.,  Mo H. J.,  van  den  Bosch F. C.,  Jing Y. P., 2005,
\mnras , 356, 1293

\bibitem[\protect\citeauthoryear{}{}]{Y06}
Yang  X., Mo  H. J., van den Bosch  F. C., Jing  Y. P., Weinmann S.
M., Meneghetti  M., 2006, MNRAS, 373, 1159

\bibitem[\protect\citeauthoryear{}{}]{Y07a}
Yang X.,  Mo H. J.,  van  den  Bosch F. C., Pasquali A., Li C.,
Barden M., 2007,  ApJ, 671, 153 (Y07)

\bibitem[\protect\citeauthoryear{}{}]{Y08a}
Yang X.,  Mo H. J.,  van  den  Bosch F. C.,  2008, preprint, (arXiv:
astro-ph/0808.0539)

\bibitem[\protect\citeauthoryear{}{}]{}
Zhao D. H., Jing Y. P., Mo H. J., B\"{o}rner  G., 2003, ApJ, 597, 9

\bibitem[\protect\citeauthoryear{}{}]{}
Zheng Z. , et al. , 2005, ApJ, 633, 791
\bibitem[\protect\citeauthoryear{}{}]{}
Zheng Z., Weinberg D. H., 2007, ApJ, 659, 1



\end{thebibliography}
\end{document}